\documentclass{IEEEtran}

\usepackage{etoolbox}
\newbool{oj}
\boolfalse{oj}

\makeatletter
\ifbool{oj}{%
  \renewcommand{\p@subsection}{\thesection-}%
  \renewcommand{\p@subsubsection}{\thesection-\thesubsection}%
}{}
\makeatother

\usepackage[nolist,nohyperlinks]{acronym}
\usepackage{algorithmic}
\usepackage{amsmath,amssymb,amsfonts}
\usepackage{booktabs}
\usepackage{cite}
\usepackage{graphicx,color}
\usepackage{siunitx}
\usepackage[caption=false,font=footnotesize]{subfig} 
\usepackage{textcomp}
\usepackage{xcolor}
\usepackage{xparse} 

\usepackage{graphicx}
\usepackage[colorlinks=true, allcolors=blue]{hyperref}

\usepackage{multirow}
\usepackage{tikz}
\usetikzlibrary{arrows.meta, positioning}

\def\BibTeX{{\rm B\kern-.05em{\sc i\kern-.025em b}\kern-.08em
    T\kern-.1667em\lower.7ex\hbox{E}\kern-.125emX}}
\AtBeginDocument{\definecolor{ojcolor}{cmyk}{0.93,0.59,0.15,0.02}}

\newcommand{\cuc}[1]{\ifbool{oj}{\MakeUppercase{#1}}{#1}}
\let\oldsection\section
\let\oldsubsection\subsection

\RenewDocumentCommand{\section}{s o m}{%
  \IfBooleanTF{#1}
    {\oldsection*{\cuc{#3}}} 
    {%
      \IfValueTF{#2}
        {\oldsection[#2]{\cuc{#3}}} 
        {\oldsection{\cuc{#3}}}     
    }%
}
\renewcommand{\subsection}[1]{\oldsubsection{\cuc{#1}}}
\ifbool{oj}{\patchcmd{\thebibliography}{\section*}{\oldsection*}{}{}}{}

\begin{document}

\bstctlcite{IEEEexample:BSTcontrol} 

\begin{acronym}
    \acro{AE}{autoencoder}
    \acro{AUC}{area under the curve}
    \acro{AWGN}{additive white Gaussian noise}
    \acro{BER}{bit error rate}
    \acro{BPSK}{binary phase-shift keying}
    \acro{CFR}{channel frequency response}
    \acro{CIR}{channel impulse response}
    \acro{CNN}{convolutional neural network}
    \acro{COTS}{commercial off-the-shelf}
    \acro{CP}{cyclic prefix}
    \acro{CSI}{channel state information}
    \acro{DCI}{downlink control information}
    \acro{DL}{downlink}
    \acro{EVM}{error vector magnitude}
    \acro{FEC}{forward error correction}
    \acro{GNN}{graph neural network}
    \acro{GNSS}{global navigation satellite system}
    \acro{IFFT}{inverse fast-Fourier transform}
    \acro{ISM}{industrial, scientific, and medical}
    \acro{JSNR}{jammer-to-signal-plus-noise ratio}
    \acro{KL}{Kullback-Leibler}
    \acro{LO}{local oscillator}
    \acro{LQI}{link quality indicator}
    \acro{LTE}{long-term evolution}
    \acro{LTX}{legitimate transmitter}
    \acro{ML}{machine learning}
    \acro{NDT}{network digital twin}
    \acro{NF}{noise figure}
    \acro{NN}{neural network}
    \acro{NPN}{non-public network}
    \acro{OCSVM}{one-class support vector machine}
    \acro{OFDM}{orthogonal frequency-division multiplexing}
    \acro{OFDMA}{orthogonal frequency-division multiple access}
    \acro{PDR}{packet delivery ratio}
    \acro{PSD}{power spectral density}
    \acro{PHY}{physical layer}
    \acro{RB}{resource block}
    \acro{RE}{resource element}
    \acro{ROC}{receiver operating characteristic}
    \acro{RSS}{received signal strength}
    \acro{RSSI}{received signal strength indicator}
    \acro{SC}{subcarrier}
    \acro{SCS}{subcarrier spacing}
    \acro{SINR}{signal-to-interference-and-noise ratio}
    \acro{SJR}{signal-to-jamming ratio}
    \acro{SNR}{signal-to-noise ratio}
    \acro{STFT}{short-time Fourier transform}
    \acro{SU}{sensing unit}
    \acro{URLLC}{ultra-reliable low-latency communication}
    \acro{VAE}{variational autoencoder}
    \acro{WGN}{white Gaussian noise}
    \acro{WiFi}{Wireless Fidelity}
\end{acronym}

\ifbool{oj}{
    \receiveddate{XX Month, XXXX}
    \reviseddate{XX Month, XXXX}
    \accepteddate{XX Month, XXXX}
    \publisheddate{XX Month, XXXX}
    \currentdate{XX November, 2026}
    \doiinfo{OJCOMS.2026.XXXXXX}
}{
}

\makeatletter
\ifbool{oj}{}{%
    \def\ps@IEEEtitlepagestyle{%
        \def\@oddhead{\hbox{}\hfil\parbox[t]{0.9\textwidth}{\centering\normalfont\small\itshape
        This work has been submitted to the IEEE for possible publication.\\ Copyright may be transferred without notice, after which this version may no longer be accessible.}%
        \hfil\llap{\@IEEEheaderstyle\thepage}}%
        \let\@evenhead\@oddhead
        \let\@oddfoot\@empty
        \let\@evenfoot\@empty}%
}
\makeatother

\title{Spectrum Anomaly Detection in OFDMA Systems: Simulation Framework and Benchmark Dataset}

\ifbool{oj}{
    \author{ANTON SCHÖSSER, MOHAMMADHADI SALEHI, SINUO MA, PHILIPP SCHULZ, AND GERHARD P. FETTWEIS \IEEEmembership{(Fellow, IEEE)} \\
    \textmd{\scriptsize Vodafone Chair for Mobile Communications Systems, Technische Universität Dresden, 01062 Dresden, Germany}
    }
    \corresp{CORRESPONDING AUTHOR: Anton Schösser (e-mail: anton.schoesser@tu-dresden.de).}
    \authornote{This work was supported by the Federal Ministry of Research, Technology and Space, Germany (BMFTR) as part of the project {6G-CampuSens} (16KISK207), and within the Clusters4Future project “SEMECO” under contract number 03ZU1210CA, as well as by the Deutsche Forschungsgemeinschaft (DFG, German Research Foundation) under contract number 578350423. The authors alone are responsible for the content of the paper.}
    \markboth{Spectrum Anomaly Detection in OFDMA Systems: Simulation Framework and Benchmark Dataset}{Schösser \textit{et al.}}
}{
    \author{
        \IEEEauthorblockN{Anton Schösser, Mohammadhadi Salehi, Sinuo Ma, Philipp Schulz, and Gerhard Fettweis}\\
        \IEEEauthorblockA{Vodafone Chair Mobile Communications Systems, Technische Universit{\"a}t Dresden, Germany\\
        \small Email: \{anton.schoesser, mohammadhadi.salehi, sinuo.ma, philipp.schulz2, gerhard.fettweis\}@tu-dresden.de}
    }
}

\ifbool{oj}{
}{
    \maketitle
}

\begin{abstract}
Wireless connectivity underpins modern society and industry, enabling critical applications such as 5G \ac{URLLC} for industrial automation. However, the openness of the wireless medium exposes it to spectrum anomalies, including unintentional interference and malicious jamming, which threaten communication and sensing functionalities in 5G and emerging 6G networks. Despite its importance, spectrum anomaly detection research is hindered by a lack of publicly available datasets reflecting real-world scenarios. To address this, we present a benchmark dataset for spectrum anomaly detection in \ac{OFDMA} systems, a core technology for 5G and beyond. The dataset includes spectrograms generated across a distributed network of sensing units, covering five distinct jammer types, from simple noise to advanced pilot-aware attacks. These anomalies are simulated in an industrial factory environment using a versatile open-source framework developed and published as part of this work, enabling extensibility to new scenarios and interference types. We provide baseline evaluations for supervised and unsupervised learning methods, demonstrating the challenges posed by different jammers and highlighting areas for further research. The dataset and framework support reproducible studies and serve as a foundation for advancing spectrum anomaly detection, with applications extending to context-aware detection that exploits side information such as resource allocation. By bridging the gap in open dataset availability, this work empowers the research community to validate and compare advanced detection methods for resilient next-generation wireless systems.
\end{abstract}


\ifbool{oj}{
    \maketitle
}{
}

\acresetall%

\section{Introduction}
\IEEEPARstart{W}{ireless} connectivity has become an integral part of everyday life and is playing a pivotal role in the ongoing automation and digitalization of industry by enabling more flexible, cost-effective manufacturing systems. In this context, the focus is increasingly shifting from traditional human-to-human communication towards human-to-machine and machine-to-machine use cases, which are addressed, for example, by 5G \ac{URLLC}~\cite{gundall2021introduction}.

At the same time, wireless networks are inherently prone to interference and other impairments that can degrade network quality, which poses a serious risk to mission- and safety-critical applications. Beyond typical interference, for instance, from coexisting technologies, the literature has discussed a variety of jammer types and related anomalies, including description~\cite{lichtmann20185g, arjoune2020smart, pirayesh2022jamming} and practical implementations~\cite{birutis2022practical, flores2023implementation, skokowski2024practical} of potential attacks. More broadly, spectrum anomalies also include unintended interference of any kind. It is important to highlight that this threat extends not only to wireless communications but also to sensing~\cite{yildrim2025ofdm}, which is expected to be an integral part of evolving 6G communication networks. Consequently, spectrum monitoring and the detection of anomalies in the observed spectrum are key capabilities for future generations of wireless networks~\cite{testi2024wireless}.

Many works have addressed the problem of detecting spectrum anomalies, considering different types of anomalies and jammers and applying miscellaneous approaches~(see Section~\ref{sec:related-work:detection}). However, suitable publicly available datasets are rare (see Section~\ref{sec:related-work:datasets}) and primarily focus on naive jammer types. This makes it challenging to quantitatively compare the different approaches that are proposed. Moreover, a relevant dataset needs to include challenging cases, such as deceptive jammers or unintended interferers that transmit legitimate-like signals (e.g., neighboring networks or devices connected to such a network). In such scenarios, jamming can be realized by generating seemingly legitimate waveforms (e.g., via MATLAB) or by recording legitimate signals and replaying them, which makes detection substantially harder. Moreover, smart jamming that exploits knowledge of specific signaling (e.g., pilots) is reported to be particularly effective while remaining difficult to detect~\cite{clancy2011efficient, lichtmann20185g}. This variety of anomalies is not explored sufficiently in the literature; instead, problem formulations are commonly treated as an either-or setting, i.e., unsupervised detection of strongly deviating patterns versus supervised classification of a small set of predefined jammer types.

To address this gap in future research, we provide a comprehensive dataset for spectrum anomaly detection in \ac{OFDMA} systems, together with the corresponding simulation framework. In detail, our major contributions are:
\begin{enumerate}
    \item \textit{Benchmark with resource information and metadata:} We provide a benchmark dataset of spectrograms of a 5G-like \ac{OFDMA} system in an industrial environment, a domain with a critical need for resilience~\cite{arendt2021pushing, michaelides2025secure}. Since the legitimate signals are simulated, their time-frequency resource allocation is controllable and published as ground truth for every sample, which, to the best of our knowledge, no existing public dataset offers. Each sample is further annotated with the jammer type, power, and location, as well as the signal-to-noise and signal-to-jammer ratio per \ac{SU}, which enables an evaluation stratified by difficulty. In addition, each sample is assigned to the training, validation, or test set, so that results obtained on this predefined split are directly comparable across studies.
    \item \textit{Applicability to 5G and 6G:} The legitimate signals follow the 5G resource grid structure, i.e., users are scheduled on resource blocks and slots. The dataset therefore captures the time-frequency structure of 5G networks and, since \ac{OFDMA} is expected to remain the basis of multiple access in 6G, is also applicable to spectrum monitoring in future networks.
    \item \textit{Jammer diversity:} The dataset covers five jammer types of increasing sophistication, from barrage to deceptive and pilot jammers, extending spectrogram-based datasets to anomalies that resemble or specifically target the legitimate signals.
    \item \textit{Distributed sensing:} Every sample is observed by 21 spatially distributed \acp{SU} through physically consistent, ray-traced channels, reflecting that a monitoring system covers an area rather than a single reception point.
    \item \textit{Simulation framework:} We make the underlying framework publicly available, so that further environments, wireless technologies, and anomaly types can be investigated.
    \item \textit{Baselines and outlook:} To demonstrate the use of the dataset, we benchmark two established detection approaches (supervised and unsupervised) across the jammer types as a reference for future work, and outline research directions such as context-aware spectrum monitoring.
\end{enumerate}

The remainder of this paper is organized as follows. Section~\ref{sec:related-work} provides an overview of detection methods and related publicly available datasets. The system model that forms the foundation of the simulation and dataset presented in this work is detailed in Section~\ref{sec:system-model}, while Section~\ref{sec:dataset} provides an in-depth discussion of the simulation framework and the structure of the published dataset, together with a recommended evaluation protocol. As a baseline, a comparative analysis of two detection approaches on the dataset is executed in Section~\ref{sec:example-use}. Recommendations for further studies utilizing the provided dataset are outlined in Section~\ref{sec:further-studies}.

\section{Related Work}
\label{sec:related-work}

The growing importance of spectrum anomaly detection, coupled with the rapid advancement of \ac{ML}, has spurred significant research activity in this domain. This section surveys recent anomaly detection approaches and reviews publicly available datasets relevant to the problem. In doing so, it highlights the absence of a benchmark dataset for \ac{OFDMA} systems that combines spectrograms of sophisticated jammer types with ground-truth resource information -- the precise gap that this work aims to fill.

\subsection{Spectrum Anomaly and Jammer Detection}
\label{sec:related-work:detection}

This section provides a selection of relevant works necessary to understand the significance of the attached dataset. Due to the wide variety of topics in this rapidly evolving field, we only present a small selection of studies here; the interested reader is referred to comprehensive surveys such as~\cite{pirayesh2022jamming} and~\cite{lohan2024from} for a more detailed exploration.

Most of the state-of-the-art works apply \ac{ML} to address the detection problem, with the major categories being supervised and unsupervised learning. Supervised learning is particularly suited for detecting signals with known characteristics, such as specific jammer types. This approach is implemented, for example, in~\cite {zhang2023detection}, which applies supervised learning to the continuous wavelet transform of the signal to classify different types of smart jamming. Similarly, applying different supervised learning algorithms, a binary classification is set up in~\cite{varotto2024detecting-super-and-unsuper, varotto2024detecting-supervised} to detect a narrowband noise jammer in a 5G system. In addition, supervised learning is also widely employed to detect jamming of \ac{GNSS} signals, typically using spectrograms as well~\cite{morales2019jammer, swinney2021gnss}.

Spectrograms are also a typical input for unsupervised approaches. In \cite{rajendran2019unsupervised, zhou2021radio, tian2022unsupervised, kim2025anomaly}, different flavors of the same underlying principle are applied. With flavors, we refer hereby to different architectures of the \ac{NN}, consideration of semi-supervised approaches, enhancements in the training procedure, etc. The underlying principle is that \acp{AE} are trained on spectrograms under normal conditions. If now spectrograms containing anomalies are presented to the \ac{AE}, the proper reconstruction fails, since anomalous patterns have not been present in the training phase. The reconstruction error, which is the difference between the input and the output, is subsequently employed as an indicator for anomalies. Another spectrogram-based approach utilizes segmentation to detect spectrum anomalies of unspecified kind~\cite{sabanovic2024ai}. In addition to the spectrogram, another approach that was recently demonstrated for jammer detection is leveraging images of the constellation diagram, together with the \ac{AE}-based approach~\cite{sciancalepore2024jamming, varotto2024one}.

Next to the \ac{ML}-based methods, other methods have been proposed recently. Those consider, for instance, the \ac{EVM} with a threshold-based approach~\cite{ornek2022an, ornek2024an}, or pseudo-random blanking of subcarriers~\cite{chiarello2021jamming}. Another line of work pursues context-aware detection, i.e., detection that exploits side information about the expected legitimate signal, such as scheduling and resource allocation or a reference model of the propagation environment. For instance, a \ac{NDT} can provide a reference for measurable metrics, such as the received power~\cite{schoesser2025advancing} or the \ac{CFR}~\cite{schoesser2025leveraging}.

\subsection{Datasets}
\label{sec:related-work:datasets}

\begin{table*}
    \centering
    \caption{Related publicly available datasets. Data level distinguishes signal-level data (IQ samples, spectrograms), power-only measurements, and link-level quality metrics reported by the receiver.}
    \label{tab:datasets}
    \ifbool{oj}{}{\renewcommand{\arraystretch}{1.2}}
    \footnotesize
    \setlength{\tabcolsep}{3pt}
    \begin{tabular}{cllllcccccccc}
    \toprule
         & & & & & \multicolumn{6}{c}{\textbf{Anomaly Types}} & \multicolumn{2}{c}{\textbf{Further Features}} \\ \cmidrule(lr){6-11} \cmidrule(l){12-13}
         \textbf{Ref.} & \textbf{Year} & \textbf{Legitimate Signals} & \textbf{Data Level} & \textbf{Data Type}
            & \shortstack{\footnotesize\bfseries Noise} & \shortstack{\footnotesize\bfseries Tone} & \shortstack{\footnotesize\bfseries Sweep/\\\footnotesize\bfseries Chirp}
            & \shortstack{\footnotesize\bfseries Pulsed/\\\footnotesize\bfseries Periodic} & \shortstack{\footnotesize\bfseries Deceptive} & \shortstack{\footnotesize\bfseries Smart/\\\footnotesize\bfseries Reactive}
            & \shortstack{\footnotesize\bfseries Multi-\\\footnotesize\bfseries Sensor} & \shortstack{\footnotesize\bfseries Resource\\\footnotesize\bfseries Allocation} \\ \midrule
         \cite{kosmanos2021rf}      & 2021          & IEEE 802.11p  & Link          & \acs{RSSI}, \acs{SINR}, \acs{PDR}, \dots      &            &            &            &            & \checkmark & \checkmark &            &            \\
         \cite{alhazbi2023dataset}  & 2023          & \acs{BPSK}    & Signal        & IQ samples                                    & \checkmark & \checkmark &            &            &            &            &            &            \\
         \cite{hussain2023jamming}  & 2023          & IEEE 802.11   & Power         & \acs{RSS}                                     & \checkmark &            &            & \checkmark &            &            &            &            \\
         \cite{kim2024wasd}         & 2024          & \acs{LTE}, 5G & Signal        & Spectrograms                                  &            & \checkmark & \checkmark & \checkmark &            &            &            &            \\
         \cite{ali2024rf}           & 2024          & IEEE 802.11   & Power         & Spectral scan (power, \dots) & \checkmark &          &            &            &            &            &            &            \\
         \cite{panitsas2025jamshield} & 2025        & IEEE 802.11   & Link          & \acs{RSSI}, \acs{SINR}, \dots                 & \checkmark &            &            &            &            & \checkmark &            &            \\ \midrule
         \textbf{Ours}              & \textbf{2026} & \textbf{5G-like \acs{OFDMA}} & \textbf{Signal} & \textbf{Spectrograms}
            & \checkmark & \checkmark & \checkmark & \checkmark & \checkmark & \checkmark & \checkmark & \checkmark \\
    \bottomrule
    \end{tabular}
\end{table*}

Most of the works related to spectrum anomaly detection -- encompassing jamming as one of its most studied forms -- and particularly the ones utilizing \ac{ML} methods, leverage datasets to train and evaluate their proposed approach. However, only a small portion of the works make their data available to the community to allow reproducibility or allow others to demonstrate enhancements over state-of-the-art works. Table~\ref{tab:datasets} enumerates the open datasets identified in this survey; their modest number reflects a broader gap between the pace of algorithmic development and the infrastructure needed to support reproducible, cross-study comparisons. 

Particularly important for spectrum anomaly detection are the works \cite{alhazbi2023dataset} and \cite{kim2024wasd}, since both IQ samples and spectrograms allow a direct and fine-grained access to the spectrum. Another approach is to consider the \ac{RSS} as done by~\cite{hussain2023jamming}, which might be suitable for deployment on small sensor nodes but can be expected to be of limited significance in dynamic scenarios. The dataset in~\cite{ali2024rf} provides spectral-scan metrics, such as noise and power values. Datasets considering jamming in IEEE~802.11 networks, in contrast, mostly report link-level metrics such as the \ac{RSSI}, \ac{SINR}, or \ac{PDR}~\cite{kosmanos2021rf, panitsas2025jamshield}. These are easily accessible from \ac{COTS} Wi-Fi chipsets. However, such metrics capture only the effect of an anomaly on the link rather than the anomalous signal itself. Consequently, a degradation cannot be unambiguously attributed to an interferer, since fading, mobility, or congestion produce similar symptoms, and the anomaly can neither be characterized nor located in time and frequency. Moreover, detection becomes possible only once the link has already been degraded. Basing the detection directly on IQ samples or spectrograms avoids these limitations and allows interferers to be detected in low-\ac{BER} regimes, i.e., before they cause a significant impact~\cite{sciancalepore2024jamming}.

Regarding the anomalies, Table~\ref{tab:datasets} shows that most datasets focus on jamming patterns that significantly deviate from legitimate signals, i.e., noise, tone, sweep, or pulsed waveforms, which makes them comparatively easy to detect. Only \cite{kosmanos2021rf} and \cite{panitsas2025jamshield} include deceptive, smart, or reactive jammers, but they provide solely link-level metrics. The closest dataset to ours is WASD~\cite{kim2024wasd}, which offers spectrograms of \ac{LTE} and 5G signals and thus also covers \ac{OFDMA} transmissions. However, its anomalies are limited to pulse, chirp, and tone patterns, and, to the best of our knowledge, it neither provides the resource allocation of the legitimate users nor covers multiple spatially distributed receivers.

The dataset presented in this work closes this gap (see last row of Table~\ref{tab:datasets}): it combines spectrograms of deceptive and pilot jammers with the ground-truth resource allocation of the legitimate users and with observations from multiple spatially distributed \acp{SU}. Moreover, each sample is annotated with the jammer type, power, and location as well as per-\ac{SU} difficulty metrics such as the \ac{SJR}, and assigned to predefined data splits. The dataset is thereby complementary to measurement-based datasets, which capture real hardware impairments but offer neither this ground truth nor control over the scenario.

\section{System Model}
\label{sec:system-model}

The system model underlying the simulation and generated dataset consists of the following parts: the general scenario, including the physical environment the network operates in, the legitimate and anomalous signals that are transmitted in this environment, and the receiver part, finally leading to the spectrograms that form the core of the presented dataset. Those are introduced in the following.

\subsection{General Scenario and Physical Environment}

\begin{figure}[t]
    \centering
    \definecolor{sceneLTX}{RGB}{0,255,88}
    \definecolor{sceneJammer}{RGB}{255,88,0}
    \definecolor{sceneSU}{RGB}{254,248,0}
    \tikzset{legmark/.style={circle, inner sep=0pt, minimum size=5pt, draw=black!50, line width=0.3pt}}
    {\footnotesize
    \tikz\node[legmark, fill=sceneLTX]{};~Legitimate transmitter\quad
    \tikz\node[legmark, fill=sceneJammer]{};~Jammer\quad
    \tikz\node[legmark, fill=sceneSU]{};~Sensing unit\par}
    \vspace{7pt}
    \includegraphics[width=\linewidth]{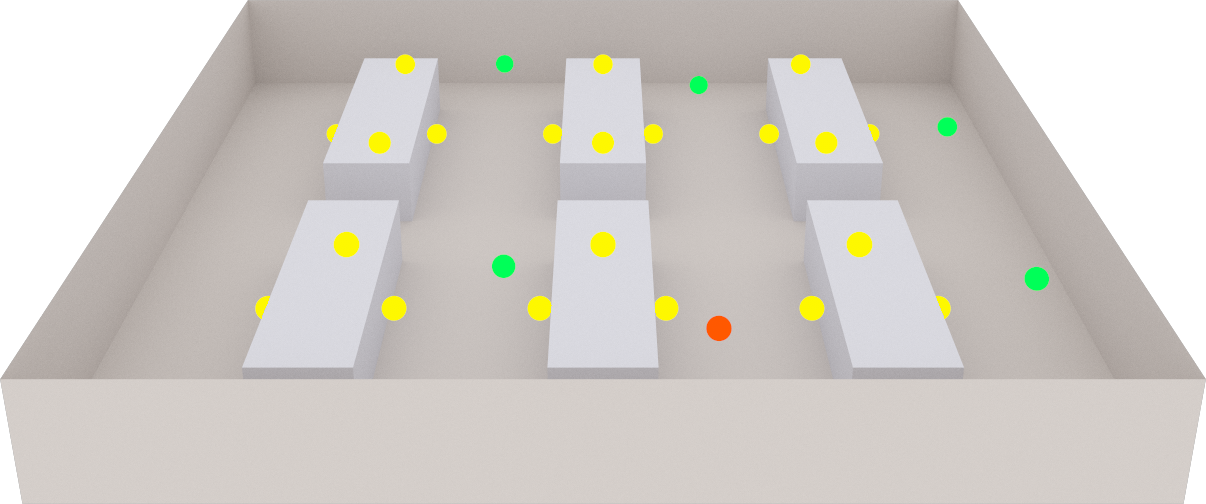}
    \caption{The simplified factory environment utilized for ray tracing simulations with five \acp{LTX} and a jammer present. The rectangular boxes represent production lines. All 21~\acp{SU} are shown: twelve attached to the sides of the production lines and nine mounted at the ceiling, which appear on top of the production lines in this view. For visual clarity, the ceiling itself is not rendered.}
    \label{fig:scene}
\end{figure}

We consider a wireless network deployed in an industrial factory environment, operating in a licensed frequency band to support the large bandwidths required by modern industrial wireless use cases~\cite{ojanen2019assessment}. This scenario is motivated by the critical importance of resilience in industrial settings, where network disruptions can lead to costly production downtimes and significant operational losses. We orient our scenario towards a \ac{NPN} deployed in a licensed band. Although licensed bands inherently offer a reduced level of interference compared to unlicensed \ac{ISM} bands, they do not provide a guarantee against accidental or intentional interfering signals. Consequently, continuous spectrum monitoring and the timely detection of all interfering signals -- treated as anomalies in the received spectrum -- constitute a key feature of robust industrial wireless networks~\cite{michaelides2025secure}.

\begin{table}[t]
\centering
\caption{Simulation parameters}
\label{tab:sim-parameters}
\ifbool{oj}{}{\renewcommand{\arraystretch}{1.2}}
\begin{tabular}{lll}
\toprule
\textbf{Parameter}                                          &                           & \textbf{Value}                                        \\
\midrule
\textbf{Geometric properties}                               &                           &                                                       \\
\hspace{0.05cm} Factory hall                                &                           & $\SI{40}{m} \times \SI{40}{m} \times \SI{5}{m}$       \\
                                                            &                           & Concrete                                              \\
\hspace{0.05cm} Production lines                            &                           & $\SI{14}{m} \times \SI{4}{m} \times \SI{3}{m}$        \\
                                                            &                           & Metal                                                 \\
\hspace{0.05cm} Number of \acp{SU}                          & $N_\text{SU}$             & 21                                                    \\
\hspace{0.3cm} at production lines                          &                           & 12, height \SI{1.5}{m}                                \\
\hspace{0.3cm} ceiling-mounted                              &                           & 9, height \SI{4.9}{m}, $3 \times 3$ grid              \\
\textbf{Radio properties}                                   &                           &                                                       \\
\hspace{0.05cm} Center frequency                            & $f_c$                     & \SI{3.75}{GHz}                                        \\
\hspace{0.05cm} System bandwidth                            & $B$                       & \SI{20}{MHz}                                          \\
\hspace{0.05cm} Total number of \acsp{SC}                   & $N_\text{SC}$             & 1320                                                  \\
\hspace{0.05cm} Number of slots                             & $N_\text{slots}$          & 5                                                     \\
\hspace{0.05cm} Number of \acsp{LTX}                        & $N_\text{TX}$             & 3, ..., 10                                               \\
\hspace{0.05cm} Transmit power of \acsp{LTX}                & $P_\text{tx, LTX}$        & \SI{10}{dBm}                                          \\
\hspace{0.05cm} Transmit power of jammer                    & $P_\text{tx, jam}$        & 0, 5, ..., 30 \si{dBm}                                 \\
\hspace{0.05cm} Antenna pattern \ac{LTX}                    &                           & isotropic                                             \\
\hspace{0.05cm} Antenna pattern jammer                      &                           & TR 38.901                                             \\
\hspace{0.05cm} Material properties                         &                           & ITU-R P.2040-3                                        \\
\bottomrule
\end{tabular}
\end{table}

To reflect on the considered industrial environment, a simplified factory building, which is shown in Fig.~\ref{fig:scene}, is used to model the physical environment. This scene is utilized for ray tracing, executed using \textit{Sionna}~\cite{sionna}, to obtain physically consistent channel realizations for a multitude of \acfp{LTX}, potentially a jammer, and distributed \acp{SU} serving as receivers. The simplified environment has an area of $\SI{40}{m} \times \SI{40}{m}$. In this area, there are six production lines, represented by metal cuboids. This layout is loosely inspired by~\cite {niu2022from}. Within this environment, a wireless network employing an \ac{OFDMA} scheme is deployed. In the time interval that is considered for detection (defined in the next paragraph), $N_\text{TX}$ \acp{LTX} are assumed to be active in the monitored system bandwidth $B$ (the \acp{LTX} are referred to with index $i \in \mathcal{T} =\{1, \ldots, N_\text{TX}\}$). They are placed at uniformly distributed random locations in the x-y-plane -- but not inside the metal cuboids -- at a fixed height of~\SI{1.5}{m} and are transmitting signals generated as described in Section~\ref{sec:system-model:legitimate-signals}. For a share of the samples, there is an additional transmitter emitting an interfering signal, referred to as a jammer. It is placed in the same way as the \acp{LTX}, but is equipped with a directional antenna. Thus, it is also assigned an orientation that is drawn randomly for each jammer.

To monitor the wireless spectrum, distributed \acp{SU} are deployed on the observed area (referred to with index $j \in \mathcal{S} =  \{1, \ldots, N_\text{SU}\}$). Twelve \acp{SU} are attached to the production lines at a height of \SI{1.5}{m}, i.e., at the same height as the transmitters, and nine are mounted at the ceiling at \SI{4.9}{m} in a $3 \times 3$ grid; their exact positions are provided with the dataset. The \acp{SU} receive the signals generated according to Sections~\ref{sec:system-model:legitimate-signals} and~\ref{sec:system-model:anomaly-signals}, and provide spectrograms which are the core part of the dataset introduced in this paper. Each spectrogram is assumed to cover a bandwidth $B$ of \SI{20}{MHz} and a time $T$ of \SI{5}{ms}, with further details motivating these parameters provided in the next section. The simulation parameters are summarized in Table~\ref{tab:sim-parameters}.

\subsection{Legitimate Signals}
\label{sec:system-model:legitimate-signals}

The resource grid underlying the simulated networks is oriented towards the 5G resource grid, motivated by the potential deployment of 5G in an \ac{NPN} setup. Hence, following the 5G terminology, 12~\acp{SC} are considered as one \ac{RB}, and $N_\text{symb}=14$ \ac{OFDM} symbols in time form one slot. Thus, the considered bandwidth~$B$ of \SI{20}{MHz} spans $N_\text{RB}=110$~\acp{RB}. One \ac{RB} per slot is the smallest resource entity assigned to one user. A \ac{SCS} of \SI{15}{kHz} is assumed. Following the typical 5G specification and considering a \ac{CP} of $\SI{4.69}{\micro s}$, the slot length is \SI{1}{ms}, and for one spectrogram we consider $N_\text{slots} = 5$ slots, leading to an observation duration per spectrogram $T$ of \SI{5}{ms}. At this stage of the work, no protocol-specific signaling is implemented, ensuring a flexible and generalized approach to spectrum analysis.

\subsubsection{Resource Allocation}

Given the orientation towards a 5G network, we assume a centralized resource allocation model as the foundation for generating legitimate users' signals. This centralized allocation ensures that no collisions occur, meaning each time-frequency resource is assigned to at most one user. However, in practice, this assumption can be violated due to various factors, such as inter-cell interference caused by imperfections in time synchronization. We treat such violations as either negligible or, if significant, as anomalies that warrant detection.

\begin{figure}
    \centering
    \includegraphics[]{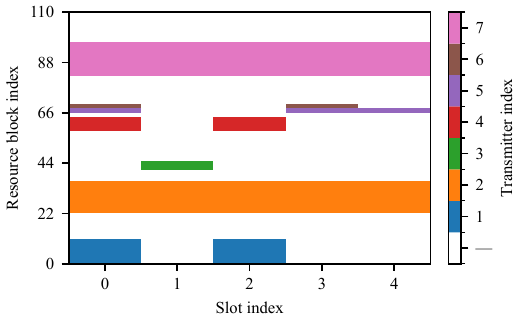}
    \caption{Example resource allocation map for $N_\text{TX}=7$ \acp{LTX}. White ("---") indicates resources that are not allocated. The map belongs to the sample whose spectrogram is shown for the deceptive jammer in Fig.~\ref{fig:example-spectrograms}(e).}
    \label{fig:resource_allocation_image}
\end{figure}

For resource allocation, we do not account for certain  traffic patterns because of the short time intervals considered; instead, we adopt the following approach. Each transmitter is assigned a contiguous block of \acp{RB} in the frequency domain, where the number of \acp{RB} per transmitter is drawn uniformly, and any remaining unallocated \acp{RB} are distributed as guard gaps between adjacent transmitters. In the time domain, each transmitter's activity is governed by a per-transmitter utilization rate $p_{\text{util},i} \sim \mathcal{U}(0.3, 1)$, with individual slots activated independently via Bernoulli draws at rate~$p_{\text{util},i}$, and a fallback mechanism guaranteeing at least one active slot per transmitter; this models bursty, heterogeneous traffic patterns that are characteristic of realistic wireless systems. The resulting assignment is summarized in the \textit{resource allocation map} of a sample, which indicates for each \ac{RB} in each slot the \ac{LTX} it is allocated to, or that it is not allocated. An example of a resource allocation map is provided in Fig.~\ref{fig:resource_allocation_image}.

\subsubsection{Signal Generation}

Based on the allocated resources, the transmit signals for the \acp{LTX} are generated. For this, the number of allocated \acp{RE} (one \ac{RE} equals one \ac{OFDM} symbol at one \ac{SC}) is determined, with which, together with the specified modulation order, a bit sequence of the required length is generated. This is used to create the modulated symbols, which are mapped to the allocated \acp{RE}. We omit the implementation of \ac{FEC}, as it does not affect the spectrograms: in systems such as 5G, the coded bits are scrambled before modulation and are thus practically independent and uniformly distributed, which corresponds to how the bit sequences are generated here. Likewise, instead of inserting specific pilot sequences, we only specify their locations within the resource grid, which is required for the pilot jammer (see next section), and fill them with data symbols. Since the spectrograms only reflect the power per \ac{RE}, this is equivalent to transmitting pilot sequences of the same average power. As usual, the \ac{OFDM} symbols are converted to the time domain using the \ac{IFFT} and the \ac{CP} is appended. In the last step, the signals are scaled to meet the specified transmit power $P_\text{tx, LTX}$. The outcome of the signal generation step is the time-domain signal of each \ac{LTX}.

\subsection{Anomaly Signals}
\label{sec:system-model:anomaly-signals}

\begin{figure*}[t]
    \centering

    \includegraphics[width=\textwidth]{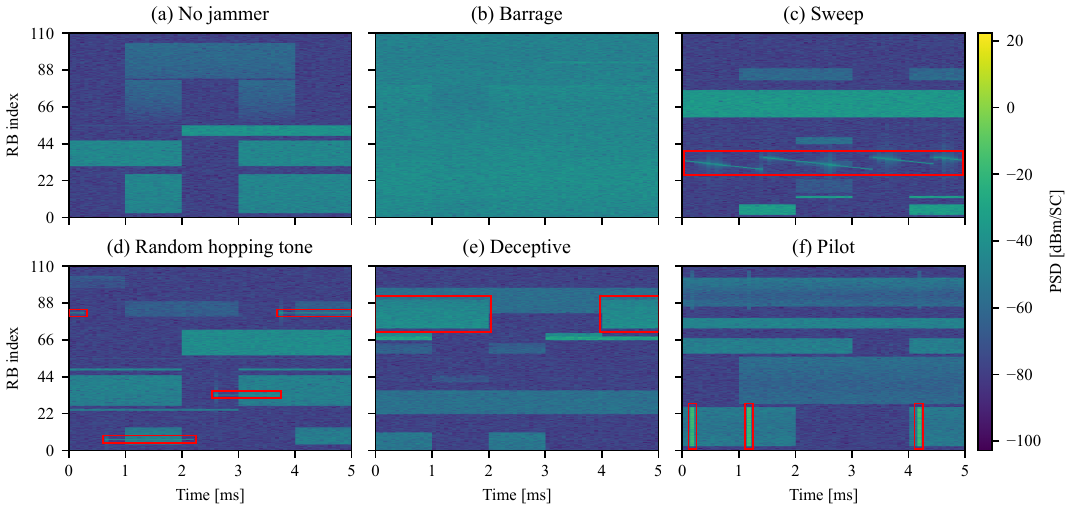}

    \caption{Example spectrograms for normal operation and the different jammer types, each at one \ac{SU}. The red boxes indicate the affected regions. Even though the spectrograms provided in the dataset are gray-scale images, we show a colored version here for improved human perception. Further details are provided with the anomaly type definitions.
    Besides the jammers, the spectrograms reveal the hardware impairments of the \acp{SU} (Sec.~\ref{sec:system-model:su-hardware}): the residual DC offset appears as a horizontal line at the center subcarrier (particularly visible in (a) and (d)), and strong signals produce an image due to IQ imbalance, mirrored at the center frequency (particularly visible in (f) for the pilot jamming).
    }
    \label{fig:example-spectrograms}
\end{figure*}

For the anomaly detection task, we implement five types of jammers that represent different characteristics and levels of system knowledge and complexity. Details on the different jammer types are provided below. For all of them, we assume a directional antenna and a transmit power $P_\text{tx, jam}$ that is drawn with equal probability from $\{0, 5, \ldots, 30\}$\,\si{dBm}. The jammer properties are also summarized in Table~\ref{tab:sim-parameters}. In all samples, there is at most one jammer, which is placed at a random location inside the factory hall at the same height as the \acp{LTX}. Fig.~\ref{fig:example-spectrograms} shows an example spectrogram for each jammer type, together with one for normal operation in Fig.~\ref{fig:example-spectrograms}(a).

\subsubsection{Barrage Jammer}

The barrage jammer is the simplest type of jammer, which emits a time-continuous \ac{AWGN} signal spanning the full system bandwidth. An example spectrogram showing the effect of the barrage jammer is provided in Fig.~\ref{fig:example-spectrograms}(b). It can be seen that even though the transmitted signal is white, different frequencies are affected differently due to the channel fading characteristics.

\subsubsection{Sweep Jammer}

The sweep jammer, exemplarily shown in Fig.~\ref{fig:example-spectrograms}(c), is characterized by four parameters, which are the instantaneous bandwidth, the total swept bandwidth, the center frequency of the swept bandwidth, and the sweeping interval. The jamming signal is a noise signal of the instantaneous bandwidth, which is randomly selected between one \ac{SCS} and 24~\acs{SCS} (i.e., two~\acp{RB}), and the center frequency is swept during the sweeping interval in a way that the swept bandwidth is covered during one sweeping interval. Sweeps are implemented with both positive and negative frequency slopes.

\subsubsection{Random Hopping Tone Jammer}

The term tone jammer refers to a jammer that emits a simple sine signal. In this work, we implement a tone jammer that is combined with frequency hopping and periodic jamming. For this, a hop interval and a duty cycle are chosen randomly. For each interval, a random frequency of the tone signal is selected. One realization of this jammer is shown in Fig.~\ref{fig:example-spectrograms}(d). For brevity, it is referred to with \textit{random hop} in the following.

\subsubsection{Deceptive Jammer}

A deceptive jammer is a jammer that reduces detectability by emitting a legitimate-like signal~\cite{pirayesh2022jamming}. Thus, it requires a certain degree of knowledge of the target system and entails greater complexity than the three previous types. For the implementation chosen in this work, we assume the jammer is aware of the underlying resource grid and, in addition, synchronized in time and frequency to the network. The jammer utilizes this information to transmit an \ac{OFDM} signal that resembles the legitimate signals by respecting the slot structure and covering random bandwidths similar to the legitimate signals that follow the \ac{RB} allocation. However, it does not react to the actual \acp{LTX}, thus it potentially (partially) overlaps with the signals of one or even more \acp{LTX}. An exemplary deceptive jamming signal is shown in Fig.~\ref{fig:example-spectrograms}(e). It can be seen that the deceptive signal looks very similar to the legitimate ones, with the difference that it overlaps, which does not happen if there are only legitimate signals. The resource allocation map of the same sample (Fig.~\ref{fig:resource_allocation_image}) reveals that the jammer occupies partly unallocated \acp{RB} and partly overlaps with \ac{LTX}~7.

\subsubsection{Pilot Jammer}

For the pilot jammer, we consider an enhanced version of the jammer that was practically demonstrated in~\cite{flores2023implementation}. In this paper, the authors showcase a jammer targeting a 5G system that decodes the \ac{DCI} and leverages it to jam the uplink resources of a specific user. We further sophisticate this approach by assuming the jammer concentrates its energy on the pilot symbols of a target user, which are assumed to be at a known location in the time-frequency grid, specifically, as implemented here in one fixed \ac{OFDM} symbol of every slot at every second \ac{SC}. This way, the jammer increases its efficiency while at the same time reducing its detectability~\cite{clancy2011efficient}. In the example in Fig.~\ref{fig:example-spectrograms}(f), the jamming signal thus appears as short high-power bursts, which occur only in the slots in which the target user is active.

\subsection{Received Signals at the Sensing Units}

\subsubsection{Channel and Noise}
\label{sec:system-model:channel-noise}

\begin{figure}
    \centering
    \includegraphics[width=.85\linewidth]{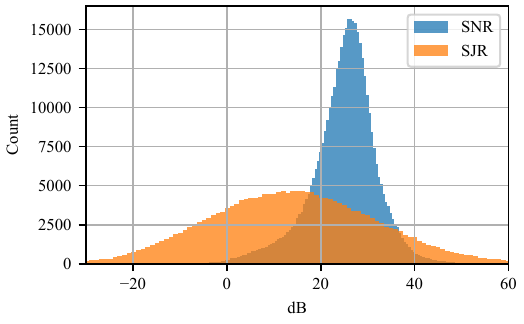}
    \caption{Empirical distribution of the \ac{SNR} and the global \ac{SJR} (both averaged over the full time-frequency grid) for all spectrograms in the dataset. For \ac{SNR}, the bin width is \SI{0.5}{dB}, whereas for the \ac{SJR} the bin width is set to \SI{1}{dB} (\ac{SJR} only applies for the anomaly samples). Since the jammer is only in the vicinity of a fraction of the \acp{SU} in each sample, a large part of the spectrograms are characterized by high \ac{SJR} (i.e., a comparably weak jamming signal).}
    \label{fig:snr_sjr_distribution}
\end{figure}

Using ray tracing, the \ac{CIR} between each \ac{LTX} (and, if present, the jammer) and the \acp{SU} is estimated. The received signals are then computed as the convolution of the time-domain signals with the corresponding \acp{CIR}. From this, the power of the received signal is determined by calculating the time-averaged power of the sum of the superimposed \ac{LTX} signals. If a jammer is present, the \ac{SJR} is computed as the ratio of the received power of the \acp{LTX} signals to the received power of the jammer signal. The distribution of the \ac{SJR} across the dataset is illustrated in Fig.~\ref{fig:snr_sjr_distribution}.

Next, the noise signal for each \ac{SU} is generated. For the noise power, we consider three components. The foundation is the thermal noise, calculated for the bandwidth $B = \SI{20}{MHz}$ and a temperature of \SI{290}{K}. Moreover, the \ac{NF} of the \acp{SU} is considered, which is set to \SI{10}{dB}. The last component is an empirical noise margin of \SI{15}{dB}, which matches the \ac{SNR} levels to commonly expected values for indoor scenarios~\cite{cebecioglu2024experimental} and avoids having only very-high \ac{SNR} spectrograms in the dataset. The hardware non-idealities of the \acp{SU} are not part of this margin, but are modeled separately (see Section~\ref{sec:system-model:su-hardware}). Subsequently, a \ac{WGN} signal is generated, scaled to the desired power, and added to the received signals. The resulting distribution of \ac{SNR} values for all spectrograms is shown in Fig.~\ref{fig:snr_sjr_distribution} (see also the comment on the \ac{SNR} and \ac{SJR} calculation below).

\textit{Comment on the \ac{SJR} and \ac{SNR} calculation:} With multiple \acp{LTX} and signals that are not necessarily active for the whole observation time, \ac{SJR} and \ac{SNR} are not uniquely defined. Here, both are global quantities, i.e., the powers of the \acp{LTX}, the jammer, and the noise are averaged over the full time-frequency grid. Consequently, both scale with the number of \acp{LTX}, and the \ac{SJR} increases if a jammer is only active for a fraction of the time (concerning pilot, random hop, and partially deceptive). Moreover, the \ac{SJR} does not reflect whether a jammer concentrates its power on a few \acp{SC}. The global \ac{SJR} hence describes the jammer's footprint in relation to the overall signals, whereas the jammer's strength within the resources it actually occupies is described by the local metrics provided in the metadata (see Section~\ref{sec:dataset:metadata}).

\subsubsection{Hardware Non-Idealities}
\label{sec:system-model:su-hardware}

In spectrum anomaly detection, hardware non-idealities are relevant since they cause spectrogram artifacts that can be mistaken for anomalies. We model them only for the \acp{SU}, since these impair every observed signal, assuming mid-grade, calibrated direct-conversion receivers that are synchronized to the network in time and frequency. The parameters are drawn uniformly from the ranges below once per \ac{SU} and kept fixed for all samples, since they are a property of the device; the drawn values are provided with the dataset. The included non-idealities are:
\begin{itemize}
    \item \textit{IQ imbalance}: a gain and phase mismatch between the two branches creates an attenuated mirror image of each received signal on the opposite side of the band center. Gain and phase mismatches of 0.02 to \SI{0.05}{dB} and $0.15^\circ$ to $0.6^\circ$ yield an image rejection ratio of 45 to \SI{55}{dB}. This exceeds the 30 to \SI{40}{dB} of careful analog design alone~\cite{valkama2001advanced} and corresponds to calibrated integrated transceivers, e.g., the typical quadrature errors of \SI{0.2}{\percent} and $0.2^\circ$ specified in~\cite{ad9361} imply about \SI{54}{dB}. The image is therefore mostly buried in the noise and becomes visible only for strong signals, e.g., for the pilot jammer in Fig.~\ref{fig:example-spectrograms}(f), which concentrates its power on few \acp{RE}.
    \item \textit{DC offset}: self-mixing of \ac{LO} leakage~\cite{razavi1997design} adds a constant to the received signal, which appears as a narrow line at the band center over the whole observation time (see, e.g., Fig.~\ref{fig:example-spectrograms}(a) and~(d)). With zero-based indexing, the band center is \ac{SC}~660. Due to windowing for the \ac{STFT}, the DC offset spans over \ac{RB}~54 and~55. The total power of the residual offset after calibration is set to 5 to \SI{20}{dB} above the noise power per frequency bin. This range is our assumption, since reported values~\cite{peng2020blind} refer to the converter full scale and depend on the receiver gain.
    \item \textit{Residual synchronization offsets}: the spectrogram time windows are offset from the \ac{OFDM} symbol boundaries by up to \SI{390}{ns} (about \SI{8}{\percent} of the \ac{CP}), the 3GPP initial transmit timing-error limit for user equipment at \SI{15}{kHz} \ac{SCS}~\cite{3gpp38133}. The \ac{LO} deviates from the network carrier by up to $\pm 0.1$\,ppm (\SI{375}{Hz} or \SI{2.5}{\percent} of the \ac{SCS}), the corresponding frequency-error limit~\cite{3gpp38101-1}. Both offsets barely change the magnitude spectrograms, as the timing offset stays within the \ac{CP} and the frequency offset causes negligible leakage into neighboring \acp{SC}.
\end{itemize}
Further effects, such as phase noise, quantization, and nonlinear distortion, are not considered.

\subsubsection{Spectrogram Computation}

Based on the final time signal including \acp{LTX}, noise, and, if applicable, the jammer's signal, the spectrograms are calculated. This is done using the \ac{STFT} with the Blackman-Harris window with a length of 2048 samples. We assume that the spectrum monitoring system is synchronized to the wireless system, up to the residual timing offset introduced above, and therefore, the hop size follows the \ac{OFDM} symbol duration in time. The final spectrograms are provided in the logarithmic domain and are denoted by $S[k,l]$, where $k$ denotes the frequency bin and $l$ the time index.

\subsection{Modeling Assumptions and Limitations}

The simulation abstracts from real-world spectrum monitoring in several respects, which we briefly discuss to clarify the validity of the dataset. Hardware non-idealities and synchronization errors are particularly relevant, since the resulting artifacts potentially challenge the detection; they are modeled for the \acp{SU} as described in Section~\ref{sec:system-model:su-hardware}. Mobility is not modeled explicitly: since the factory environment is not characterized by high mobility, neither large Doppler shifts nor large channel variations are expected within the observation time of \SI{5}{ms}. On longer timescales, mobility is reflected by the varying placement of the \acp{LTX} across samples. Similarly, traffic patterns play only a minor role within the short observation time, such that a dedicated traffic model would add little realism compared to the randomized resource allocation described in Section~\ref{sec:system-model:legitimate-signals}. Finally, each anomalous sample contains a single jammer, as multiple simultaneous jammers add considerable complexity to the evaluation and the interpretation of the results.

Since we publish the code underlying this simulation framework, the scenario and all related aspects can be flexibly adjusted, for instance, to consider a different environment, other communication technologies, further anomaly types, or multiple simultaneous jammers.
\section{Simulation Framework and Dataset}
\label{sec:dataset}

The presented system model has been implemented in a simulation framework that has been made publicly available, together with the generated dataset. This section provides a brief overview of the steps that are taken in the simulation and of the dataset that is provided.

\subsection{Simulation Framework}

The simulation framework introduced in this paper is implemented in Python, utilizes Blender for scene generation, and Sionna~\cite{sionna} for executing ray tracing to compute the channels within the scene. The full code is available on \textit{GitHub}\footnote{\url{https://github.com/akdd11/ofdma-spectrum-anomalies-simulation}}. Generating a dataset using the provided framework consists of three steps that are introduced in the following; more details can be found in the corresponding repository.

\subsubsection{Scene Generation}

To run ray tracing, a scene needs to be created that defines both the geometry and materials of objects in the considered environment. Several options are available for this. Either the scene described in Section~\ref{sec:system-model} or other scenes already defined in Sionna can be used. Moreover, own scenes can be created in Blender, which is facilitated for outdoor scenarios by the \textit{BLOSM addon for Blender}, or for simple indoor scenes by the script provided in our repository.

\subsubsection{Configuration}

Before running the simulation, the simulation parameters need to be configured, which is done via a configuration file. This concerns parameters such as bandwidth and center frequency, number of \ac{LTX}, resource grid structure,~etc.

\subsubsection{Data Generation}

The final step involves running the simulation, which produces an intermediate data format. This intermediate format is subsequently transformed into images and a CSV file containing metadata for each sample. The metadata can serve multiple purposes, such as providing labels, acting as additional input data for training, or further structuring the evaluation process.

\subsection{Dataset}

The provided dataset consists of \num{20000} samples in total, whereby each sample comprises the spectrogram for each \ac{SU} as well as the metadata. Moreover, the resource allocation map of each sample is provided in an image format as well. The dataset is available on \textit{Zenodo}~\cite{dataset}. Details on the available data types are provided in the following.

\subsubsection{Spectrogram Images}

The spectrograms are saved as 8-bit grayscale PNG files with a resolution of $N_\text{SC} \times (N_\text{slots} \cdot N_\text{symb})$ and contain the \ac{PSD} in a logarithmic scale. Prior to saving the images, they are clipped to a dataset-wide range and min-max scaled to 8 bit. The lower bound is the 0.1st percentile of all values, so that the sparse tail far below the noise floor does not waste gray levels. The upper bound is the maximum value of all non-jammed samples plus a margin of \SI{6}{dB}, so that no normal spectrogram is clipped. Note that this scaling is a prerequisite for storing the data as PNG files and does not replace the normalization that is required for a \ac{ML} workflow (see Section~\ref{sec:dataset:evaluation-protocol}). The clipping bounds and information on rescaling can be found in the data description.

\subsubsection{Resource Allocation Maps}

The resource allocation map of each sample is stored as an 8-bit grayscale PNG image, where each pixel value corresponds to the index of the \ac{LTX} to which the resource is assigned (with 0 indicating unallocated resources). The resolution of this image is defined the same as for the spectrograms, i.e., $N_\text{SC} \times (N_\text{slots} \cdot N_\text{symb})$. This approach aims to provide additional information that can be utilized to enhance anomaly detection in licensed bands. An example of such a resource allocation map is shown in Fig.~\ref{fig:resource_allocation_image}. It is important to note, however, that the figure presented in the paper has been visually enhanced using a discrete colormap for better interpretability, rather than displaying the original grayscale values. In Section~\ref{sec:further-studies}, it is further discussed how those data can be exploited.

\subsubsection{Metadata}
\label{sec:dataset:metadata}

The provided metadata contains both labels, i.e., the jammer type, as well as additional information for each sample. The features are summarized in Table~\ref{tab:metadata}, together with their column names in the CSV file.

\begin{table}
    \centering
    \caption{Metadata provided for every sample. The features per \ac{SU} are stored in one column for each \ac{SU} index $j$.}
    \ifbool{oj}{}{\renewcommand{\arraystretch}{1.2}}
    \label{tab:metadata}
    \setlength{\tabcolsep}{4pt}
    \newcommand{\col}[1]{{\scriptsize\texttt{#1}}}
    \begin{tabular}{@{}lll@{}}
    \toprule
      \textbf{Feature}                       & \textbf{Symbol}                & \textbf{Column name}                   \\ \midrule
      \textbf{Per sample}                    &                                &                                        \\
      \hspace{0.05cm} Jammer type            &                                & \col{jammer\_type}                     \\
      \hspace{0.05cm} Jammer power (\si{dBm}) & $P_\text{tx, jam}$            & \col{jammer\_power}                    \\
      \hspace{0.05cm} Jammer location (\si{m}) &                              & \col{jammer\_}$\{x,y,z\}$              \\
      \hspace{0.05cm} Number of \acp{LTX}    & $N_\text{TX}$                  & \col{num\_legitimate\_transmitters}    \\
      \hspace{0.05cm} Jammer occupancy       & $\rho$                         & \col{jammer\_occupancy}                \\
      \hspace{0.05cm} Split, supervised      &                                & \col{split\_supervised}                \\
      \hspace{0.05cm} Split, unsupervised    &                                & \col{split\_unsupervised}              \\
      \textbf{Per \acs{SU}}                  &                                &                                        \\
      \hspace{0.05cm} \acs{SNR} (\si{dB})    & $\mathrm{SNR}_j$               & \col{snr\_by\_su\_}$j$                 \\
      \hspace{0.05cm} Global \acs{SJR} (\si{dB})  & $\mathrm{SJR}_j$          & \col{sjr\_by\_su\_}$j$                 \\
      \hspace{0.05cm} Local \acs{JSNR} (\si{dB})  & $\mathrm{JSNR}_{\text{loc},j}$ & \col{jsnr\_local\_by\_su\_}$j$    \\
      \hspace{0.05cm} Global contrast (\si{dB})   & $\Delta_{\text{glob},j}$  & \col{db\_contrast\_global\_by\_su\_}$j$\\
      \hspace{0.05cm} Local contrast (\si{dB})    & $\Delta_{\text{loc},j}$   & \col{db\_contrast\_local\_by\_su\_}$j$ \\ \bottomrule
    \end{tabular}
\end{table}

The \ac{SNR} and the global \ac{SJR} have been introduced in Section~\ref{sec:system-model:channel-noise}. Since both are averaged over the full time-frequency grid, they are complemented by features that are restricted to the set $\mathcal{J}$ of \acp{RE} affected by the jammer. The jammer occupancy $\rho$ is the fraction of all \acp{RE} that belong to $\mathcal{J}$. The local \ac{JSNR} at \ac{SU}~$j$ is given by
\begin{equation}
    \mathrm{JSNR}_{\text{loc},j} = 10 \log_{10} \frac{P_{\text{jam},j}}{P_{\text{LTX},j} + P_{\text{n},j}} \; ,
\end{equation}
where $P_{\text{jam},j}$ and $P_{\text{LTX},j}$ denote the received power of the jammer and the \acp{LTX} averaged over $\mathcal{J}$, and $P_{\text{n},j}$ the noise power per \ac{RE}. The jammer power is placed in the numerator, since the \ac{LTX} power vanishes if the jammer occupies \acp{RE} that are not allocated to any \ac{LTX}, such that a ratio with the \ac{LTX} power in the numerator would not reflect the jammer's strength. Hence, in contrast to the \ac{SJR}, high values indicate a strong jammer. Moreover, the contrast
\begin{equation}
    \Delta_j[k,l] = S_j[k,l] - \tilde{S}_j[k,l]
\end{equation}
is the change in dB that the jammer causes in the spectrogram $S_j[k,l]$ of \ac{SU}~$j$, where $\tilde{S}_j[k,l]$ denotes the same spectrogram without the jammer's signal. Its mean over all \acp{RE} and over $\mathcal{J}$ is provided as global contrast $\Delta_{\text{glob},j}$ and local contrast $\Delta_{\text{loc},j}$, respectively. For samples without a jammer, the jammer-related features are empty and the \ac{SJR} is infinite.

\subsection{Recommended Evaluation Protocol}
\label{sec:dataset:evaluation-protocol}

A sample is labeled as anomalous if a jammer is present, irrespective of whether the jammer is perceptible at every \ac{SU}, and as normal otherwise. Moreover, the jammer type can serve as label for classification tasks.

To allow a consistent evaluation of different detection methods, a predefined split for both supervised and unsupervised learning is provided in the metadata columns \texttt{split\_supervised} and \texttt{split\_unsupervised}, assigning each sample to the training, validation, or test set.
While the unsupervised training and validation sets contain only normal samples, the corresponding supervised sets include both normal and anomalous samples of all jammer types. The test set is shared between both approaches and contains normal samples as well as samples from all jammer types. Samples not included in a particular split remain unassigned. The number of samples in each split is summarized in Table~\ref{tab:data-split}.
To assess the generalization to unseen jammer types, supervised methods should additionally be evaluated in a leave-one-jammer-type-out setting, i.e., for each jammer type, a separate model is trained with the samples of this type removed from the supervised training and validation sets and evaluated on this type.

To prevent data leakage, it is essential to ensure that normalization statistics, e.g., mean and variance, or minimum and maximum, are computed on the training or validation set, but never on the test set. If min-max normalization is to be applied, it is recommended to reserve an additional margin beyond the value range of the training set, in particular on the maximum values, since jammed samples may exhibit higher values than the training samples due to the presence of the jammer. This holds for both supervised and unsupervised methods, but is particularly relevant for the latter, since their normalization statistics are computed on normal samples only.

For evaluation, the \ac{AUC} is recommended as the primary metric, as it is independent of the threshold and provides a comprehensive measure of the model's performance across different operating points. All metrics are computed separately for each jammer type, based on the test samples of this type and all normal test samples.
\section{Example Use}
\label{sec:example-use}

The presented dataset facilitates advancements in spectrum anomaly detection methods by leveraging spectrograms, with a particular emphasis on multi-\ac{SU} scenarios. To illustrate the practical application of the dataset and establish a baseline for future research, we conduct a comparative analysis of two common approaches in anomaly detection: supervised and unsupervised learning.

These baselines are intended to characterize the dataset rather than to compete for state-of-the-art detection performance. Specifically, they serve to show that (i) the dataset is learnable but not trivial, (ii) the detection difficulty varies across jammer types, and (iii) known and unseen jammer types behave differently under supervised and unsupervised detection. The two methods were selected according to the following criteria. First, they represent two common paradigms in spectrogram-based detection, namely supervised classifiers and reconstruction-based \acp{AE} (cf. Section~\ref{sec:related-work:detection} and~\cite{varotto2024detecting-super-and-unsuper, rajendran2019unsupervised, zhou2021radio, tian2022unsupervised, kim2025anomaly, sciancalepore2024jamming}), which anchors the baselines in established practice. Second, their behavior is well understood, so that performance differences can be attributed to the properties of the dataset rather than to the detector. Third, both rely on standard, widely used architectures (ResNet-18 and a convolutional \ac{VAE}) and can be reproduced from the published code, configuration, data splits, and trained weights.

When comparing supervised and unsupervised learning, unsupervised approaches are considered more suitable than supervised learning methods for unconstrained anomaly detection, justified mainly by the inherently unpredictable and evolving characteristics of the unseen anomalies. On the other hand, supervised learning has been reported to efficiently detect and classify interfering signals with known characteristics~\cite{robinson2023narrowband}.

Following the evaluation protocol described in Section~\ref{sec:dataset:evaluation-protocol}, the unsupervised model is trained exclusively on normal data, whereas the supervised model is trained both on all jammer types, providing an upper-bound reference, and in the leave-one-jammer-type-out setting, such that both approaches can be compared on jammer types absent from their training sets.

\begin{figure*}[t]
    \centering
    \begin{tikzpicture}[
        font=\small\sffamily,
        node distance=0.7cm and 1cm,
        base/.style={
            draw,
            rectangle,
            rounded corners=3pt,
            minimum width=0.5cm,
            minimum height=0.5cm,
            align=center,
            line width=0.6pt
        },
        input/.style={
            draw,
            rectangle,
            rounded corners=3pt,
            minimum width=2.5cm,
            minimum height=1.2cm,
            align=center,
            line width=0.6pt
        },
        op/.style={
            circle,
            draw,
            minimum size=5mm,
            inner sep=0pt
        },
        inputstyle/.style={input, fill=blue!10},
        encoderstyle/.style={base, fill=green!10},
        latentstyle1/.style={base, fill=orange!15},
        latentstyle2/.style={base, fill=orange!55},
        latentstyle3/.style={base},
        decoderstyle/.style={base, fill=green!10},
        lossstyle/.style={base, fill=red!10},
        arrow/.style={->, line width=0.8pt},
        dashedarrow/.style={->, dashed, line width=0.7pt},
        cdashedarrow/.style={->, dashed, line width=0.7pt, draw=blue!90}
    ]
    
    \node (input) [inputstyle] {\textbf{Input}\\$x$};
    \node[anchor=south west, inner sep=0] (img) [below=0.1cm of input]
        {\includegraphics[width=0.07\linewidth]{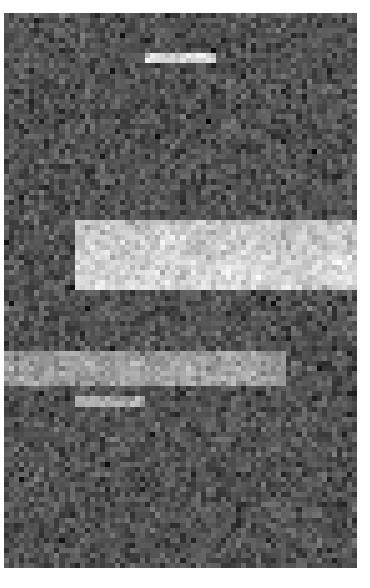}};
    
    \node (enc) [encoderstyle, right=of input] 
    {\textbf{CNN Encoder}\\$q_\phi(z|x)$};
    
    \node (z) [latentstyle2, right=1.5cm of enc] 
    {\textbf{Latent variable}\\$z=\mu+\sigma\cdot\epsilon$};
    
    \node (mu) [latentstyle1, above right=0.5cm of enc] {$\mu(x)$};
    \node (sigma) [latentstyle1, below right=0.5cm of enc] {$\sigma(x)$};
    
    \node (dec) [decoderstyle, right=of z] 
    {\textbf{CNN Decoder}\\$p_\theta(x|z)$};

    \node (cross) [op, below=0.4cm of z] {$\times$};
    \node (eps) [latentstyle3, below right=1cm of cross] {$\epsilon \sim \mathcal{N}(0,I)$};
    
    \node (output) [inputstyle, right=of dec] 
    {\textbf{Reconstruction}\\$\hat{x}$};

    \node[anchor=south west, inner sep=0] (img) [below=0.1cm of output]
        {\includegraphics[width=0.07\linewidth]{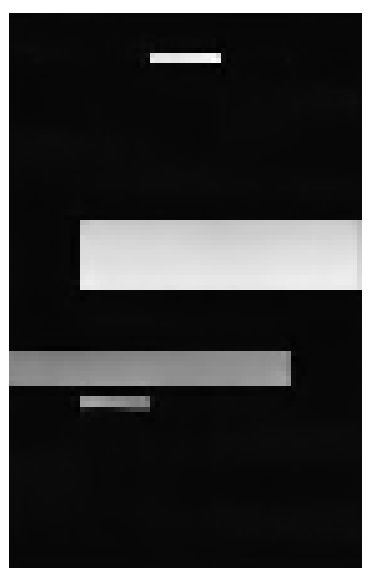}};
    
    
    
    
    \draw[arrow] (input) -- (enc);
    \draw[arrow] (enc) -- (mu);
    \draw[arrow] (enc) -- (sigma);
    
    \draw[arrow] (mu) -| (z);
    \draw[arrow] (sigma) -- (cross);
    \draw[arrow] (cross) -- (z);
    \draw[arrow] (eps) -- (cross);
    
    \draw[arrow] (z) -- (dec);
    \draw[arrow] (dec) -- (output);
    
    
    
    
    \end{tikzpicture}
    
    \caption{The model structure of the \ac{VAE}.}
    \label{fig:VAE_architecture}
\end{figure*}

\subsection{Detection Methods}

\subsubsection{Supervised Binary Classification}
\label{subsec:SupervisedLearning}

By treating anomalous and normal samples as two distinct classes, the problem can be formulated as a binary classification task within a supervised learning framework. However, as previously noted, each sample consists of spectrograms from all 21~\acp{SU}, which are not equally affected by the presence of a jammer. In an anomalous sample, the spectrograms of certain \acp{SU} may still appear consistent with normal operating conditions (e.g., SU15 in Fig.~\ref{fig:detected_anomalies}). For those \acp{SU}, the jammer can be interpreted as a \emph{hidden node}. Due to severe path loss or link obstruction, its signal arrives below the noise floor, rendering those \acp{SU} effectively unaware of the jamming. This poses a fundamental challenge for supervised learning, since a single global label cannot be consistently assigned to all \ac{SU} spectrograms within a sample. In particular, forcing an anomalous label onto spectrograms that are indistinguishable from normal behavior risks introducing label noise that degrades training.

To mitigate this inconsistency, we design a two-stage model whose first stage is supervised with per-\ac{SU} targets instead of a single global label. The first submodel, based on the ResNet-18 architecture~\cite{ResNet}, receives the spectrograms of all 21~\acp{SU} as separate input channels and predicts the global \ac{SJR} of each \ac{SU} (Table~\ref{tab:metadata}). An \ac{SU} that is barely affected by the jammer is thus assigned a high \ac{SJR} instead of an anomalous label. For jammer-free samples, whose \ac{SJR} is infinite, the target is clipped to the largest finite \ac{SJR} in the training set. The second submodel, a fully connected neural network with two layers, takes the vector of the 21 predicted \acp{SJR} as input and classifies the sample as either anomalous or normal. Note that the per-\ac{SU} \ac{SJR} targets are required only for training. While they are readily available in simulation, they are hard to obtain from real measurements, where the training can instead rely on controlled campaigns with a known jammer, transfer from simulated data, or multiple-instance learning (see Section~\ref{subsec:multi-su}).

\subsubsection{Unsupervised Anomaly Detection}
\label{subsec:VAE}

For the unsupervised learning approach to anomaly detection, we adopt a \ac{VAE} trained exclusively on normal samples.
\Acp{VAE} \cite{kingma2022autoencodingvariationalbayes} are probabilistic generative models that learn to encode data into a structured latent probability space, and decode samples from this space to reconstruct the original input while enforcing regularization of the latent representation (Fig. \ref{fig:VAE_architecture}). A \ac{VAE} assumes that observed data~$x$, i.e., the spectrogram, is generated from a latent variable $z$, where $z \sim p(z)$. Under this assumption, the encoder tries to approximate the true posterior $p(z|x)$ with a variational distribution $q_\phi(z|x)$. In the model shown in Fig. \ref{fig:VAE_architecture}, this distribution is characterized by the two parameters $\mu$ and $\sigma$, which represent the encoder's output. 
Then, a latent sample is obtained by drawing $\epsilon \sim \mathcal{N}(0,I)$ and computing $z=\mu+\sigma \cdot \epsilon$. Given the latent variable ($z$), the decoder learns to reconstruct the original input by modeling $p_\theta(x|z)$.
Both the encoder and decoder are neural networks trained jointly using a loss function composed of two terms: a reconstruction loss and a \ac{KL} divergence term. While the first term encourages accurate reconstruction of the original input, the second term penalizes deviations of the approximate posterior $q_\phi(z|x)$ from a predefined prior distribution $p(z)$.
 
\begin{figure}[tp]
\centering

    \begin{tikzpicture}
        \node[anchor=south west, inner sep=0] (img) at (0,0)
            {\includegraphics[width=0.9\linewidth]{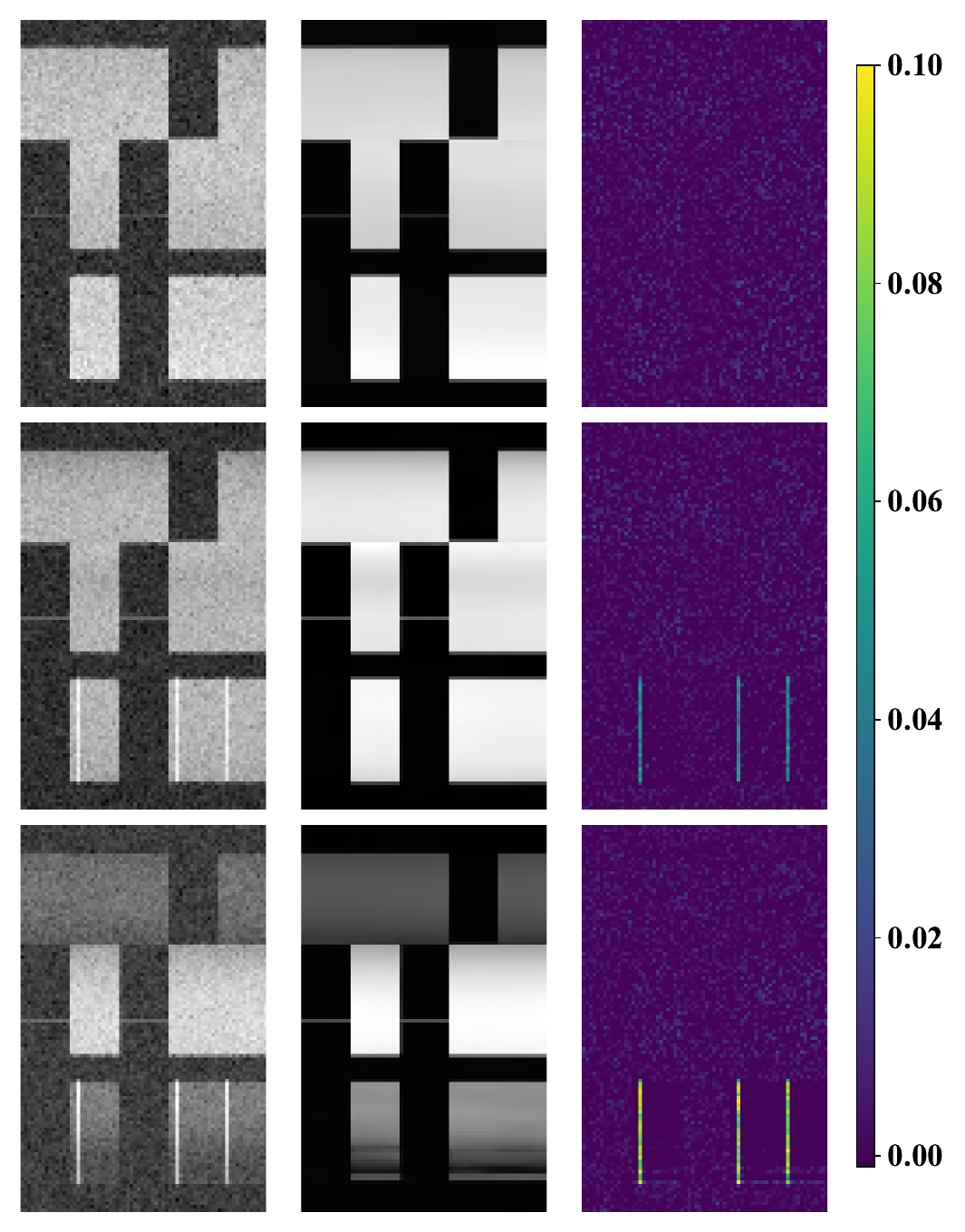}};
        
        \node[overlay, above left=0.01cm of img]
            {\begin{tikzpicture}[font=\footnotesize]
    \def\x{0.4}

    \draw[->, thick] (-0.2*\x,0) -- (\x,0); 
    \draw[->, thick] (0,-0.2*\x) -- (0,\x); 

    \node[below right] at (\x,0) {$t$};
    \node[above left] at (0,\x) {$f$};
\end{tikzpicture}};
        
        \begin{scope}[x={(img.south east)}, y={(img.north west)}]
            
            \node at (0.15,1.05) {\small Input};
            \node at (0.44,1.05) {\small \shortstack{VAE\\Reconstruction}};
            \node at (0.73,1.05) {\small \shortstack{Reconstruction \\ Error}};
            \node[rotate=90] at (-0.06,0.86) {\small SU 15};
            \node[rotate=90] at (-0.06,0.52) {\small SU 17};
            \node[rotate=90] at (-0.06,0.18) {\small SU 20};
            
        \end{scope}
    \end{tikzpicture}

    \caption{An example of three \acp{SU} of a sample with pilot jamming: only \acp{SU}~17 and 20 reveal visible impact of the jammer. 
    The \ac{VAE} removes both noise and jamming parts during reconstruction. Therefore, the reconstruction error includes both noise and jamming patterns. 
    Applying a low-pass filter to the reconstruction error images helps distinguish stochastic noise from structured jamming patterns. 
    The faint line at the center frequency is the DC offset of the \acp{SU} and therefore not a jammer signal.
    }
    \label{fig:detected_anomalies}
\end{figure}

 If we train a \ac{VAE} only on the normal data, then for any latent representation, its decoder has only learned to generate outputs that look like normal data. Therefore, abnormal areas can be detected and localized by calculating the reconstruction error for any new sample (Fig. \ref{fig:detected_anomalies}). Owing to this property, \acp{VAE} have been successfully applied to various anomaly detection tasks \cite{tian2022unsupervised,AD-VAE-Brain}.

 It should be emphasized that the KL divergence term in \acp{VAE} acts as a regularizer that enforces a smooth and structured latent space. This constraint limits the amount of information encoded in the latent variables, which encourages the model to capture meaningful structure rather than noise, often resulting in denoising behavior (Figs.~\ref{fig:VAE_architecture} and~\ref{fig:detected_anomalies}). This poses a challenge for detecting the anomalies since the reconstruction error includes both noise and the abnormalities. To address this issue, we propose applying a low-pass filter to the reconstruction error images to improve the distinction between stochastic noise and structured jamming patterns.

In this study, reconstruction error is used as the sole anomaly metric. Accordingly, the anomaly score for each sample is computed as follows:

\begin{equation}
\label{Eq: Rec_Score}
a_\text{r} = 
\max_{j,k,l} 
\left( x_j[k,l] - \hat{x}_j[k,l] \right) \;  ,
\end{equation} where $x$ is the original spectrum image and $\hat{x}$ is the \ac{VAE}'s reconstruction of the same image. Note that in this setting, we do not rely on absolute error measures such as mean squared error, since we are particularly looking for patterns that exist in the original images but are removed in the reconstruction.

\begin{figure}[tp]
\centering

\includegraphics[width=0.94\linewidth]{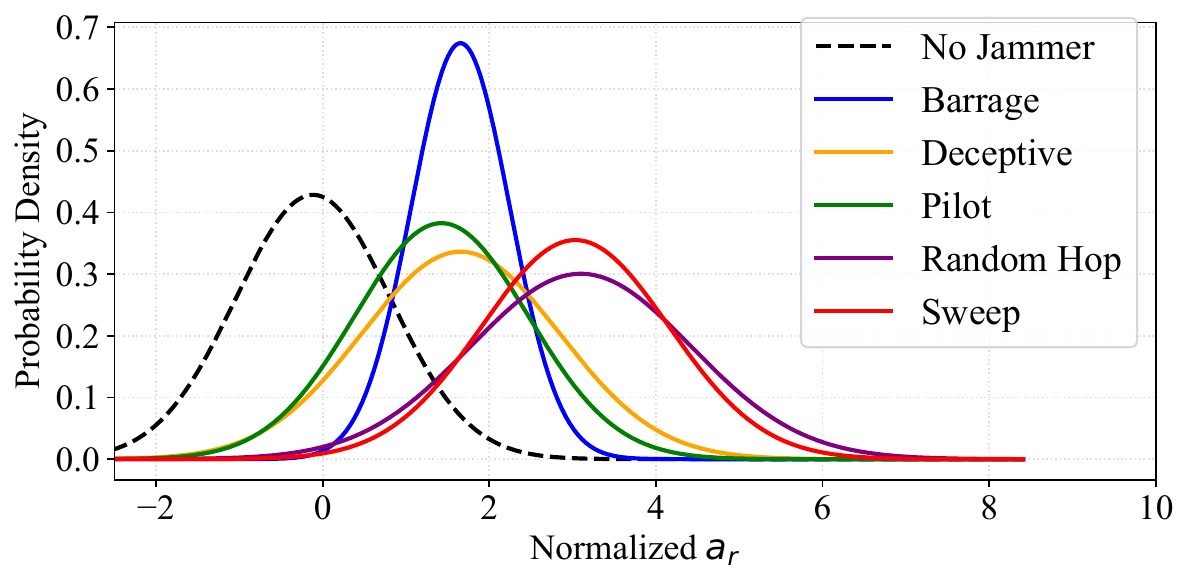}
    
\caption{Empirical distributions of \ac{VAE} anomaly score based on reconstruction error $a_{\text{r}}$ for five jammer types.}
\label{fig:VAEScores}
\end{figure}

Fig. \ref{fig:VAEScores} illustrates the distributions of the \ac{VAE}'s anomaly score for anomalous samples (colored) in comparison to normal samples (black dashed line). 

In addition to providing an anomaly score, the \ac{VAE}'s reconstruction error enables the localization of anomalies in both the time and frequency domains for jammer types that affect localized regions of the spectrum images (Fig.~\ref{fig:detected_anomalies}), as observed in earlier studies~\cite{rajendran2019unsupervised}.

\begin{figure*}[tp]
    \centering
    \includegraphics[width=0.98\textwidth]{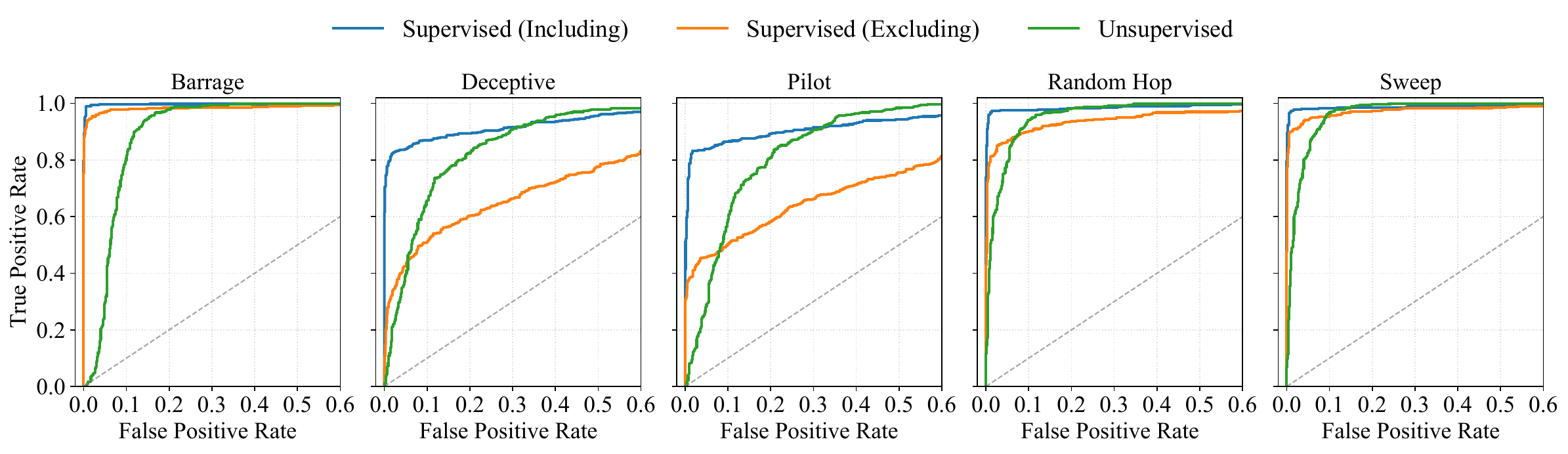}
    \caption{ROC curves comparing the performance of different anomaly detection methods on our dataset. The blue curves represent the performance of the supervised model when samples of the corresponding jammer type are included in the training set. The orange curves show its performance when the corresponding jammer type is excluded from the training set. The green curves represent the performance of the unsupervised model.}
    \label{fig:Two_Approach_Comparison}
\end{figure*}

\subsubsection{Implementation Details}

Both models are trained and evaluated following the protocol described in Section~\ref{sec:dataset:evaluation-protocol}, based on the corresponding supervised and unsupervised data splits presented in Table~\ref{tab:data-split}.
Both models are implemented in PyTorch~\cite{pytorch}. For the unsupervised approach, we designed the encoder and decoder architectures of the \ac{VAE} using residual convolutional blocks. For the supervised approach, the regression submodel is trained first, using a mean squared error loss on the \acp{SJR} min-max normalized over the training set. Its parameters are then frozen while the classification submodel is trained on its output.
To reduce the impact of noise and also the computational complexity, we reduce the image resolution by aggregating the power per \ac{RB}. Within each \ac{RB}, the signal power of the subcarriers is summed in the linear domain and subsequently converted back to the logarithmic scale, resulting in a resolution of $N_\text{RB} \times (N_\text{slots} \cdot N_\text{symb})$ for the utilized spectrograms that preserves the total energy within each \ac{RB}.
For both models, the resulting spectrograms are then min-max normalized such that the minimum and maximum values of the respective training set are mapped to 0.1 and 0.9, respectively, which leaves a margin for test samples exceeding the training range, as recommended in Section~\ref{sec:dataset:evaluation-protocol}.
The complete Python implementation, Jupyter notebooks used to generate the results, and additional information are available in the GitHub repository.

Besides the \ac{AUC} as primary metric, we report the F1 score, which requires a decision threshold. For the VAE model, anomaly scores are obtained and normalized using the mean and variance computed from the normal samples in the validation set. The distributions of these normalized scores for different jammer types in the test set are illustrated in Fig. \ref{fig:VAEScores}. The detection threshold is set to the 90th percentile of the anomaly scores obtained from the validation set. A sample is classified as anomalous if its anomaly score exceeds this threshold. The F1 score is then computed based on the resulting predictions.

For the supervised classification models, a sigmoid function is applied to the final output score to obtain the predicted probability of a sample being anomalous. These probabilities are then used to construct the ROC curves. Samples with predicted anomaly probabilities exceeding 0.5 are classified as anomalous.

\begin{table}[tp]
\centering
\caption{Distribution of samples in training, validation, and test sets}
\begin{tabular}{lcccccc}
\toprule
 & \multicolumn{3}{c}{Supervised} & \multicolumn{3}{c}{Unsupervised} \\
\cmidrule(lr){2-4} \cmidrule(lr){5-7}
 & Train & Valid & Test & Train & Valid & Test \\
\midrule
No Jammer     & 5900 & 100 & 1000 & 8800 & 200 & 1000 \\
\midrule
Barrage     & 1475 & 25 & 500 & - & - & 500 \\
Deceptive   & 1475 & 25 & 500 & - & - & 500 \\
Pilot       & 1475 & 25 & 500 & - & - & 500 \\
Random Hop  & 1475 & 25 & 500 & - & - & 500 \\
Sweep       & 1475 & 25 & 500 & - & - & 500 \\
\bottomrule
\end{tabular}
\label{tab:data-split}
\end{table}

\subsubsection{Further Detection Methods} In addition, we evaluated a per-pixel threshold-based (energy) detector and a \ac{OCSVM}. As both achieved a clearly lower \ac{AUC} than the \ac{VAE} for most jammer types, we do not report them here. Code and results are available in the repository.

\subsection{Results}
\label{sec:Detection_Results}

This section provides initial insights into detection performances with the provided dataset, based on \ac{ROC} curves, the \ac{AUC}, and the F1 score.

\subsubsection{Detection Performance on Different Jammer Types}

The detection results of both supervised and unsupervised learning approaches are presented in Fig.~\ref{fig:Two_Approach_Comparison} and Table~\ref{tab:performance_comparison}. The supervised model trained on all jammer types (i.e., the \textit{including} model) serves as a reference for assessing performance degradation when samples of a specific jammer type are absent from the training set. Based on this reference, we compare the performance of the supervised model trained without the corresponding jammer type (i.e., the \textit{excluding} model) with that of our unsupervised model based on \ac{VAE}.

The sweep and random hop jammers are both narrowband, change their frequency over time, and are not aligned with the \ac{RB} and slot structure of the legitimate signals. If either is excluded from the training set, the supervised model can rely on the patterns learned from the other, resulting in only a minor performance degradation (Fig.~\ref{fig:Two_Approach_Comparison}). The unsupervised model achieves comparable performance for both jammer types.

In contrast, pilot and deceptive jammers follow the time-frequency grid of the legitimate signals and differ from them mainly by overlapping with allocated resources. As no other jammer type shares these characteristics, the supervised model degrades strongly when they are excluded from training. The \ac{VAE} is considerably more robust, as it models normal spectrograms rather than specific jammer patterns, although these jammer types remain its most challenging ones.

Finally, barrage jammers are generally easier to detect because they affect large regions of the spectrum. This is also reflected in the strong performance of the supervised model on this jammer type, even when barrage samples are excluded from the training set. However, detecting barrage jammers using the \ac{VAE}-based model is more challenging. As discussed earlier, the reconstruction process is expected to suppress both anomalous patterns and stochastic noise. Consequently, distinguishing reconstruction errors caused by ordinary noise from those associated with barrage jammers becomes difficult, as these jammers generate noise-like interference without noticeable structure. For this jammer type, we observed that using the mean reconstruction error as the anomaly score generally yields better detection performance than using the maximum reconstruction error. However, relying on the mean reconstruction error may reduce sensitivity to localized, structured anomalies introduced by jammer types such as pilot and sweep jamming.

\begin{table}[t]
\centering
\caption{Numerical comparison of supervised and unsupervised detection.}
\resizebox{\columnwidth}{!}{
\begin{tabular}{lcccccccc}
\toprule
& \multicolumn{2}{c}{Supervised (Including)}
& \multicolumn{2}{c}{Supervised (Excluding)}
& \multicolumn{2}{c}{Unsupervised}  \\

\cmidrule(lr){2-3} \cmidrule(lr){4-5} \cmidrule(lr){6-7}
 & F1 score & AUC & F1 score & AUC & F1 score & AUC\\
\midrule
Barrage & 0.973 & 0.997 & \textbf{0.944} & \textbf{0.989} & 0.837 & 0.926 \\
Deceptive & 0.882 & 0.940 & 0.440 & 0.751 & \textbf{0.740} & \textbf{0.886} \\
Pilot & 0.884 & 0.935 & 0.530 & 0.747 & \textbf{0.705} & \textbf{0.879} \\
Random Hop & 0.962 & 0.990 & \textbf{0.882} & 0.954 & 0.872 & \textbf{0.968} \\
Sweep & 0.964 & 0.991 & \textbf{0.926} & \textbf{0.980} & 0.884 & 0.973 \\
\bottomrule
\end{tabular}}
\label{tab:performance_comparison}
\end{table}

\subsubsection{Impact of the Jammer Power on the Detection Performance}

\begin{figure*}[tp]
    \centering
    \includegraphics[width=0.96\textwidth]{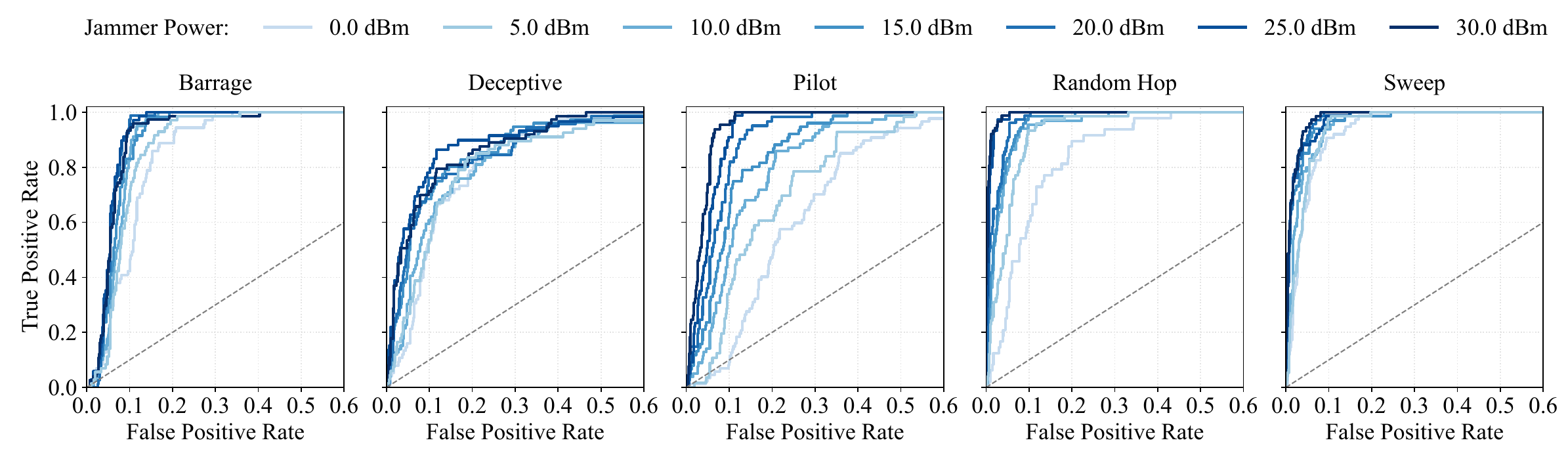}
    
    \caption{
    Dependence of the detection performance of the unsupervised model on the  transmit power of the jammer for each jammer type.
    }
    \label{fig:Power_Vs_Accuracy}
\end{figure*}

Higher jammer power is generally expected to result in more distinguishable anomalies and, consequently, higher detection accuracy. The \ac{ROC} curves of the \ac{VAE} in Fig.~\ref{fig:Power_Vs_Accuracy} illustrate how the detection performance varies with jammer power for each jammer type. For all jammer types except deceptive jamming, the \ac{ROC} curves consistently shift towards the upper left corner with the jammer power. For the deceptive jammer, in contrast, the detection performance hardly improves with increasing power, as a stronger legitimate-like signal still resembles a legitimate one. This underlines the challenge that such signals pose for spectrogram-based anomaly detection methods and motivates context-aware detection methods that, for instance, check the observed spectrum against the resource allocation (see Section~\ref{subsec:resource-allocation}).

\section{Recommendations for Further Studies}
\label{sec:further-studies}

Beyond the baselines in Section~\ref{sec:example-use}, we outline three research directions that can be pursued with the released dataset, complemented, where applicable, by extensions of the simulation framework.

\subsection{Multi-SU Inference and Jammer Localization}
\label{subsec:multi-su}

Each sample contains the spectrograms of 21~\acp{SU}, whose fixed positions are provided with the dataset, and the \acp{SU} are affected very differently by a jammer depending on its location (cf. Fig.~\ref{fig:detected_anomalies}). As each sample carries a single label and it is unknown in practice which \acp{SU} perceive the jammer (cf. the hidden node in Section~\ref{subsec:SupervisedLearning}), the detection can be formulated as a multiple-instance learning problem~\cite{carbonneau2018multiple}, with the spectrograms of all \acp{SU} forming a bag of instances. Thresholding the local \ac{JSNR} or local contrast yields instance labels, which allow quantifying the benefit of instance-level supervision. \Acp{GNN}~\cite{herzalla2026graph} offer a natural way to exploit the spatial structure of the bag: each \ac{SU} is represented by a node whose features are an embedding of its spectrogram, and the edges encode the spatial proximity of the \acp{SU}. Through message passing, the observations of neighboring \acp{SU} are combined, such that weak evidence of a jammer at a few adjacent \acp{SU} is mutually reinforced, while a max or attention readout prevents it from being diluted by the unaffected \acp{SU}.

Beyond detection, the same graph lends itself to localizing the jammer, which is an essential step towards countermeasures, as it allows the malicious device to be identified. The unknown waveform, transmit power, and antenna pattern of the jammer complicate model-based approaches such as ranging based on the received power. Instead, the jammer location provided in the metadata enables a supervised regression at the graph level~\cite{herzalla2026graph}, for which the positions of the \acp{SU} can be included in the node features. As the requirements are less strict than for industrial localization use cases, an accuracy of a few meters already narrows down the search for the malicious device. The local \ac{JSNR} or local contrast can serve as an auxiliary target at the node level, since they describe how strongly each \ac{SU} observes the jammer on the jammed \acp{RE} and thereby depend on the propagation from the jammer to the \acp{SU}. Robustness to failing \acp{SU} can be assessed by removing nodes at test time, whereas the transfer to other deployments requires data with different numbers and layouts of \acp{SU}, which the simulation framework can generate.

\subsection{Foundation Models}
\label{subsec:foundation-models}

Foundation models are pretrained on large amounts of unlabeled data and can then be fine-tuned to a specific task with few labeled samples. They are increasingly explored for wireless signals~\cite{kim2026lwmspectro,aboulfotouh2025wavesfm,liu2025spectrumfm}, where the proposed models differ in their input data, such as spectrograms, \ac{CSI} data, or IQ samples, and in the tasks they have been evaluated on, which mostly rely on the observations of a single receiver. Such models typically build on transformer architectures and are pretrained in a self-supervised manner, e.g., by reconstructing masked parts of the input or by contrastive learning. The dataset therefore allows benchmarking how well these models transfer to jammer detection from multiple distributed \acp{SU} in an \ac{OFDMA} system. Since all samples are labeled, the share of labeled training data can be reduced systematically, i.e., in a semi-supervised setting, to determine how many labeled samples a fine-tuned foundation model requires to match the supervised baseline trained on the full training set. Moreover, the leave-one-jammer-type-out setting tests whether pretrained representations generalize better to unseen anomalies, provided that the pretraining data contain neither test samples nor samples of the held-out jammer type. A pretrained model can also serve as the node encoder of the \ac{GNN}.

Besides the training spectrograms of the released dataset, the simulation framework can generate diverse data for domain-specific pretraining by varying, e.g., the scene, the number of \acp{LTX}, and the hardware non-idealities. Exporting the IQ samples, which are already computed during the simulation, requires only minimal effort and enables models that operate directly on the received signals. However, the transfer of models pretrained on simulated data to real measurements remains to be validated.

\subsection{Context-Aware Detection}
\label{subsec:resource-allocation}

The resource allocation map provided by the scheduler is particularly valuable for detecting deceptive jammers, whose signals resemble legitimate ones but ignore the allocation, such that they occupy unallocated resources or overlap with the allocation of an \ac{LTX}. Since each \ac{RB} and slot corresponds to a fixed block of \acp{SC} and symbols, the map can be upsampled to the spectrogram resolution, converted into an occupancy mask or a one-hot encoding of the \acp{LTX}, and provided as additional input channels, for instance, to the \ac{VAE}. This way, detectors can compare the observed spectrum with the expected occupancy and check the power within each allocation for consistency.

\section{Conclusion}
\label{sec:conclusion}

With the growing reliance on wireless technologies -- particularly in industrial deployments -- resilience mechanisms are increasingly essential. Because wireless operates over an open medium, systems remain vulnerable to both unintentional interference and deliberate jamming, including legitimate-like and signaling-aware attacks. This motivates continuous spectrum monitoring and anomaly detection, with relevance not only to reliable communications but also to emerging integrated sensing-and-communication use cases.

Although spectrum anomaly detection has received significant attention, reproducibility and fair comparison are limited by the shortage of public datasets, especially for challenging anomaly classes beyond simple noise-like interference. To address this gap, we provide (i) a simulation framework for an \ac{OFDMA} system with distributed \acp{SU} and (ii) a publicly available benchmark dataset spanning multiple jammer types across different complexity classes, along with the associated code. Using this benchmark, we establish supervised and unsupervised baselines that highlight a key practical trade-off: supervised models perform strongly on known interference types, yet generalization to previously unseen interferers remains challenging, motivating approaches that improve robustness to distribution shift. In this regard, context-aware detection, which leverages side information such as the resource allocation provided with the dataset, represents a promising direction, particularly for legitimate-like interference. Overall, this work enables standardized evaluation for spectrum anomaly detection and supports future research on distributed spectrum sensing and robust detection under various interference conditions.


\ifbool{oj}{\section*{ACKNOWLEDGMENT}}{\oldsection*{Acknowledgment}}

\ifbool{oj}{}{\small This work was supported by the Federal Ministry of Research, Technology and Space, Germany (BMFTR) as part of the project {6G-CampuSens} (16KISK207), and within the Clusters4Future project “SEMECO” under contract number 03ZU1210CA, as well as by the Deutsche Forschungsgemeinschaft (DFG, German Research Foundation) under contract number 578350423. The authors alone are responsible for the content of the paper.} 

\ifbool{oj}{}{\small} The authors gratefully acknowledge the computing time made available to them on the high-performance computer at the NHR Center of TU Dresden. This center is jointly supported by the Federal Ministry of Research, Technology and Space of Germany and the state governments participating in the NHR (www.nhr-verein.de/unsere-partner).

\bibliographystyle{IEEEtran}
\bibliography{database}

\ifbool{oj}{\begin{IEEEbiography}[{\includegraphics[width=1in,height=1.25in,clip,keepaspectratio]{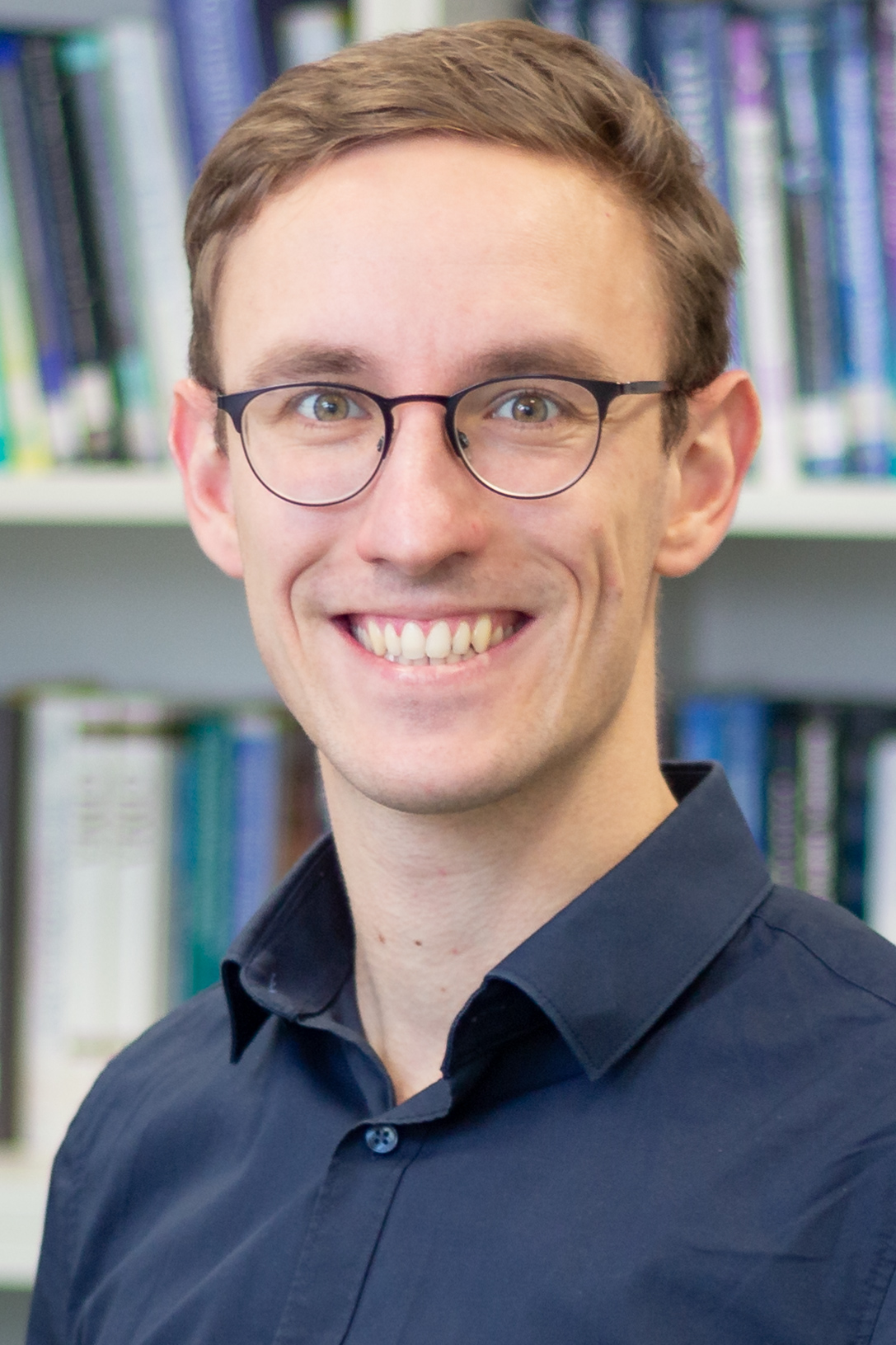}}]{ANTON SCHÖSSER } 
received his Dipl.-Ing. degree in electrical engineering in 2022 from TU Dresden, and is currently pursuing his Ph.D. at the Vodafone Chair Mobile Communications Systems at TU Dresden. His research interests include spectrum monitoring and anomaly detection, network digital twins, and radio localization.
\end{IEEEbiography}

\begin{IEEEbiography}[{\includegraphics[width=1in,height=1.25in,clip,keepaspectratio]{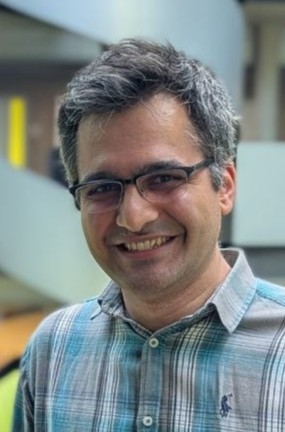}}]{MOHAMMADHADI SALEHI } received his B.Sc. degree in Electrical Engineering from Shiraz University and his M.Sc. degree in Computational Modeling and Simulation from TU Dresden, where he worked as a student research assistant at the Vodafone Chair Mobile Communications Systems. 
His research interests include multimodal machine learning, stochastic modeling, and generative AI.
\end{IEEEbiography}

\begin{IEEEbiography}[{\includegraphics[width=1in,height=1.25in,clip,keepaspectratio]{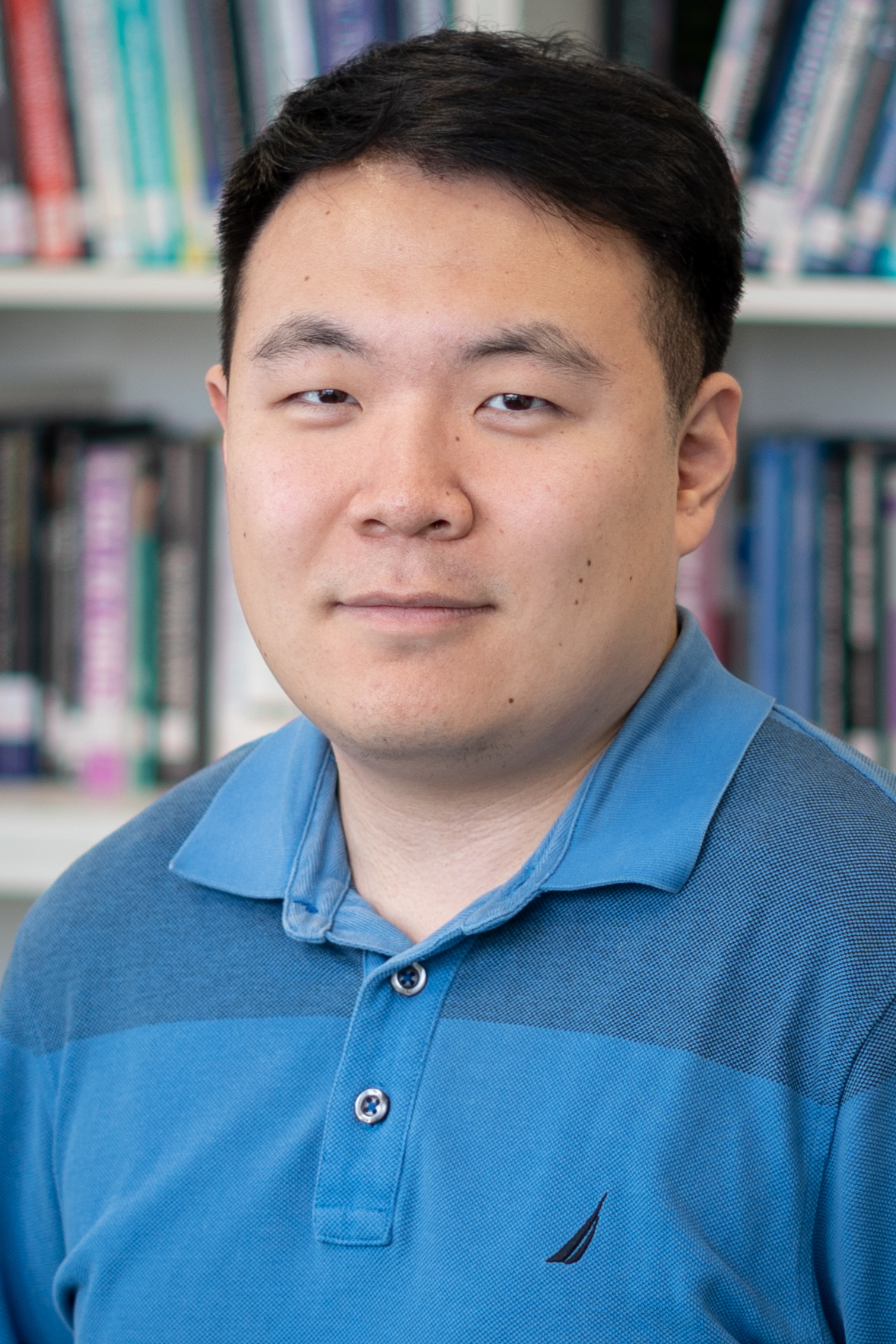}}]{SINUO MA } 
received his Dipl.-Ing. degree in electrical engineering in 2022 from TU Dresden, and is currently pursuing his Ph.D. at the Vodafone Chair Mobile Communications Systems at TU Dresden. His research interests include communication signal processing and ML for spectrum sensing.
\end{IEEEbiography}

\begin{IEEEbiography}[{\includegraphics[width=1in,height=1.25in,clip,keepaspectratio]{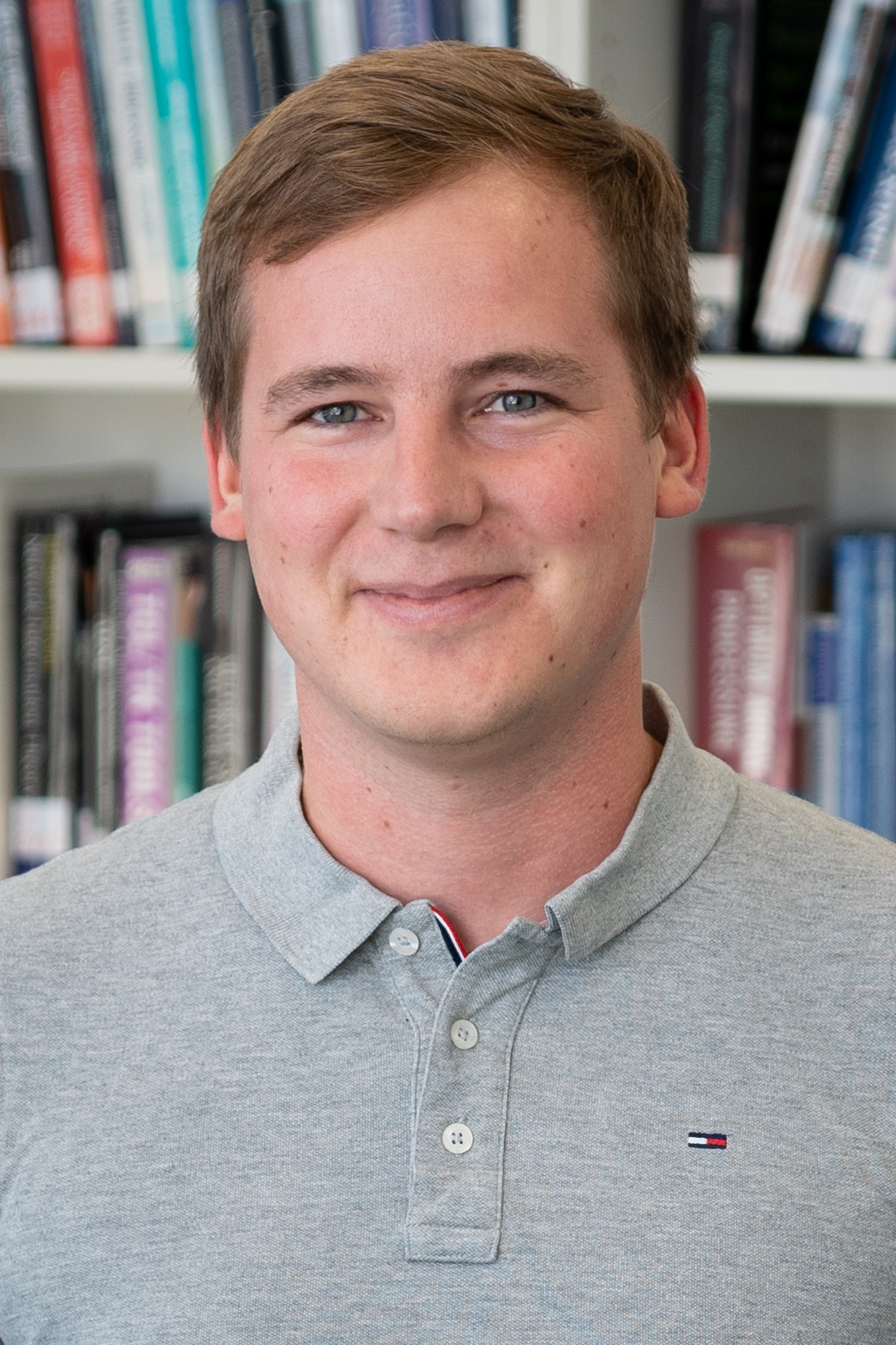}}]{PHILIPP SCHULZ } received the M.Sc. degree in mathematics and the Ph.D. (Dr.-Ing.) in electrical engineering from Technische Universität Dresden (TU Dresden), Germany, in 2014 and 2020, respectively. He was a research assistant at TU Dresden in numerical mathematics, modeling, and simulation. In 2015, he joined the Vodafone Chair for Mobile Communications Systems at TU Dresden and became a member of the system-level group. His research there focused on flow-level modeling and the application of queuing theory to communications systems for ultra-reliable low-latency communications.
After more than one year at the Barkhausen Institut in Dresden, Germany, where he investigated rateless codes in the context of multi-connectivity, he is now back at the Vodafone Chair, where he leads a research group focused on the resilience of wireless communications systems.
\end{IEEEbiography}

\begin{IEEEbiography}[{\includegraphics[width=1in,height=1.25in,clip,keepaspectratio]{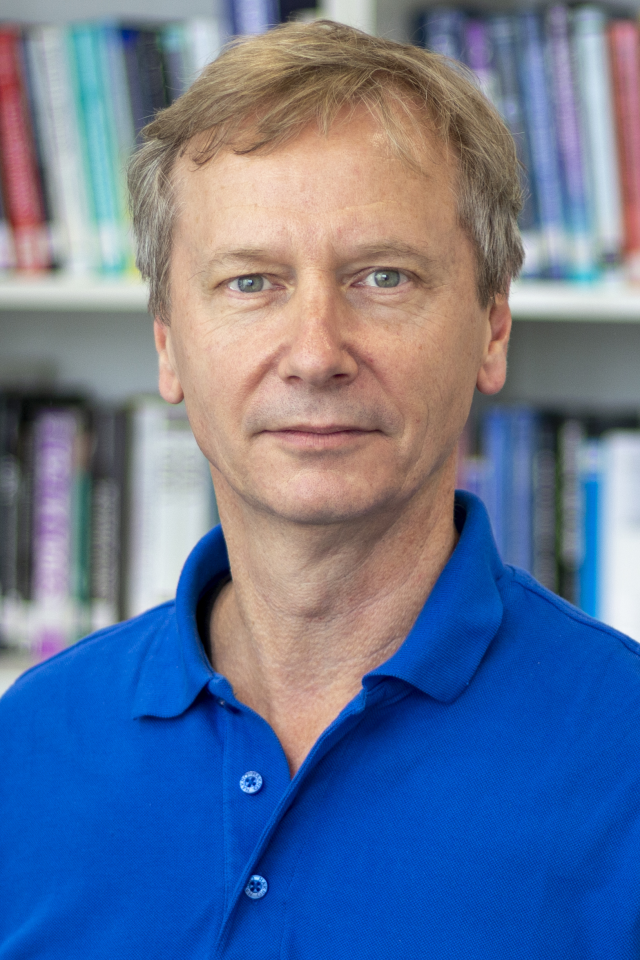}}]{GERHARD P. FETTWEIS }
(Fellow, IEEE) received the Ph.D. degree under the supervision of H. Meyr from RWTH Aachen University, Germany, in 1990. After Post-Doctoral with IBM Research, San Jose, he joined TCSI, Berkeley, USA. Since 1994, he has been the Vodafone Chair Professor with TU Dresden, Germany. Since 2018, he has been a founding Scientific Director and the CEO of Barkhausen Institute. He researches wireless communications and chip design, coordinates 5G++Lab Germany and German Cluster-for-Future SEMECO. His team spun-out 28 tech startups. He initiated six platform entities. He is a member of US National Academy of Engineering, German Academy of Sciences (Leopoldina), German Academy of Engineering (Acatech), and VDE/ITG, National Academy of Inventors, EURASIP, WWRF, and DATE. He is active in organizing IEEE conferences.
\end{IEEEbiography}}{}

\end{document}